\documentstyle[epsf]{mn}
\begin{document}       

\title[Proton synchrotron radiation]
{Proton synchrotron radiation of  
large-scale jets \\ in active galactic nuclei}
\author[F.A. Aharonian]
{F.A. Aharonian\thanks{E-mail: Felix.Aharonian@mpi-hd.mpg.de}\\
Max Planck Institut f\"ur Kernphysik,
Postfach 103980, D-69029 Heidelberg, Germany}
\date{Accepted -----     . Received -------    }
 
\maketitle                    
\begin{abstract}  
                                                          
The X-radiation of large-scale 
extragalactic jets  poses serious challenge 
for conventional electron-synchrotron or  
inverse Compton models suggested to explain 
the overall nonthermal emission of the resolved 
knots and hot spots. In this paper I propose an 
alternative mechanism for X-ray emission - 
synchrotron radiation by extremely high energy 
protons - and  discuss implications of this model 
for the extended jet features  resolved by Chandra  
in several prominent radiogalaxies and AGN 
like Pictor A, 3C 120, PKS 0637-752, and 3C 273.  
I show that if protons are indeed accelerated to energies    
$E_{\rm p} \geq 10^{18} \, \rm eV$, it is possible
to construct a realistic model which allows an effective  
cooling of protons via synchrotron radiation on quite 
``comfortable'' timescales of about $10^7 - 10^8$~yr, 
i.e. on timescales which provide effective propagation of 
protons over the jet structures on kpc scales. 
This explains quite naturally  
the diffuse  character of the observed  X-ray emission,
as well as the broad range of spectral X-ray indices 
observed from different objects. Yet, as long as 
the proton synchrotron cooling time is comparable 
with both the particle escape time and the age of 
the jet, the proton-synchrotron model offers an 
adequate radiation efficiency. The model requires 
relatively large magnetic field of about 1 mG, and 
proton acceleration rates ranging from 
$L_{\rm p} \sim 10^{43}$ to $10^{46}$
erg/s. These numbers could be reduced significantly 
if the jet structures are moving 
relativistically towards the observer. I discuss 
also possible contributions of synchrotron radiation 
by secondary electrons  produced at interactions of 
relatively low energy ($E_{\rm p} \leq 10^{13} \ \rm eV$) 
protons with the  compressed gas in the jet structures. 
This is an interesting possibility
which however requires a very large product of the 
ambient gas density and total amount of accelerated protons. 
Therefore it could be treated as a viable working 
hypothesis only if one can reduce the intrinsic X-ray 
luminosities assuming that the regions of nonthermal 
radiation are relativistically moving  condensations with  
Doppler factors $\delta_{\rm j} \gg 1$.  The kpc scales of 
knots and hot spots  are not sufficient for effective  
confinement of $\geq 10^{19} \ \rm eV$ protons.  
This suppresses the synchrotron radiation by 
secondary electrons produced at
$p \gamma$ interactions. At the same time the 
runaway protons interacting with 2.7~K 
cosmic microwave background  radiation,  
initiate non-negligible diffuse X- and 
$\gamma$-ray emission in the surrounding 
cluster environments.  I discuss the spectral 
and angular  characteristics  of this radiation  
which essentially depend  on the   
strength of the ambient  magnetic field. 
\end{abstract}

\section{INTRODUCTION}   
The recent Chandra observations revealed various  bright 
X-ray  features like knots and hot spots  in large-scale
extragalactic jets  (for a review see e.g.  Harris 2001).    
The  undisputed synchrotron origin of  the  radio 
emission  of the resolved jet structures in several 
powerful AGN and radiogalaxies implies presence 
of relativistic  electrons with a  typical Lorentz factor 
$\gamma_{\rm e} \simeq 10^{3} \ B_{\rm mG}^{-1/2}  
\nu_{\rm GHz}^{1/2}$,  
where $B_{\rm mG}=B/10^{-3} \, \rm G$ is the 
magnetic field  and $\nu_{\rm GHz}=\nu/1 \ \rm GHz$ 
is the frequency of synchrotron radiation.  
The inverse  Compton  scattering of same electrons 
leads to  the second component of nonthermal radiation. 
The 2.7 K cosmic microwave background  radiation (CMBR)  
provides a universal photon target field for the 
inverse Compton emission. The electrons 
boost  the energy  of the seed photons to  
$E \simeq 4 kT  \gamma_{\rm e}^2$; this gives 
a simple relation between the energies of inverse 
Compton and synchrotron photons produced by the 
same electron:  
$E \simeq 1 \ \nu_{\rm GHz} B_{\rm mG}^{-1}  \ \rm keV$.
Thus, for the characteristic magnetic field, which in the 
knots and hot spots is believed to be 
between 0.1~mG and 1~mG (e.g. Harris 2001), 
the inverse Compton radiation of radioelectrons 
appears in the Chandra energy domain. Yet, 
the energy density of CMBR is too low to provide 
significant  X-radiation. Indeed, the inverse 
Compton  luminosity constitutes  only a small part 
of the emission of  radioelectrons, 
$L_{\rm X} \approx 10^{-5} B_{\rm mG}^{-2} L_{\rm R}$, 
unless we assume that the radio emitting 
regions are relativistically moving condensations 
with the Lorentz factor of bulk motion 
$\Gamma \geq 10$, thus the energy density of 
CMBR ``seen'' by electrons in the jet's frame  
is enhanced by a factor of $\Gamma^2$
(Tavecchio et al. 2000; Celotti et al. 2001). 
This hypothesis automatically implies that the jet
features are moving towards the observer, otherwise 
we would  face unacceptably large requirements
to the energy budget of the central source. 

The emissivity of inverse Compton  radiation 
could be significantly enhanced  also due to  
additional seed photon sources, in particular
due to  the jet's own synchrotron radiation. 
The density of this radiation in a  knot at a distance 
$d$ from the observer is estimated  as 
$w_{\rm r}=(d/R)^2 c^{-1} f_{\rm R-O} \simeq 0.9  
\ f_{-12}  \theta^{-2} \ \rm eV/cm^3$,  where  
$f_{-12}=f_{\rm R-O}/10^{-12} \ \rm erg/cm^2 s$ 
is the observed radio-to-optical flux normalized to 
$10^{-12} \ \rm erg/cm^2 s$,  $R$ is the source 
radius and $\theta=R/d$ is the source angular 
size in arcseconds. 
For example, the  fluxes of 1 to 1000 GHz 
radio synchrotron emission of two  distinct  
hot spots  (``A'' and ``D'') 
of the  radiogalaxy Cygnus A,  both of   
$\sim 0.^{\prime \prime}5$  angular size,  
are very high,    $f \sim 10^{-11} \ \rm erg/cm^2 s$   
(Wilson et al. 2000),  thus  the density of synchrotron 
radiation  exceeds the  density of  CMBR  by two orders   
of magnitude.  Therefore,  the X-ray emission of  these  
hot spots   can be readily explained   by the  
synchrotron-self-Compton (SSC)  model, as it was predicted by 
Harris et al. (1994), assuming $B \sim  0.2$ mG, and 
taking into account that the energy flux at radio 
frequencies exceeds the X-ray flux by 1.5 orders of magnitude. 

This  is not  the case, however, for the majority 
of  jet features resolved by  Chandra 
for which we observe, in fact, just an opposite  
picture. The X-ray fluxes,  for example, from the 
jets in   Pictor A (Wilson et al. 2001), 3C 120  
(Harris et al. 1999), 3C 273 (R\"oser et al.  2000;  
Sambruna et al.  2001; Marshall et al.  2001), and  
PKS 0637-752  (Schwartz et al. 2000; Chartas et al. 2000)
are at least factor of 10 larger than the 
radio and optical fluxes.

For X-ray emission of many resolved 
extragalactic knots and hot spots  the 
electron synchrotron radiation is considered 
as `the process of choice' (Harris 2001). 
The synchrotron radiation is indeed  
an extremely effective mechanism converting  
the kinetic energy of relativistic electrons
into X-rays with an almost 100 per cent efficiency, 
provided, of course, that the electrons are accelerated to 
multi-TeV energies. The single power-law spectrum smoothly 
connecting the radio,  optical and X-ray 
fluxes with spectral index $\alpha \simeq 0.76 \pm 0.02$  
observed  from  the knot ``A1'' in the jet of 3C 273  
(R\"oser et al. 2000;  Marshall et al. 2000) 
formally could be brought as an additional argument 
for the synchrotron origin of X-rays.  
Note, however,  that the observed 
spectral steepening (or cutoff) at  
optical frequencies (but with yet 
hard X-ray  spectra) in the 
so-called ``problem sources''        
(Harris 2001) like the knot D/H3 in the 
same 3C 273 (R\"oser et al. 2000, Marshall et al. 2000),  
as well as  the knots in 3C 120 (Harris et al. 1999), 
PKS 0637-752 (Schwartz et al 2000)
and in the hot spot of  Pictor A (Wilson et al. 2000),  
tell us that we deal with a rather complex 
picture compared, in particular,  with the  
simple  {\em  single-power-law  synchrotron model}.  

Generally, the high radiation efficiency is treated as 
a key  component  of any  successful  model/mechanism   
of luminous  nonthermal sources. However, for
the large-scale extragalactic jets this highly 
desired feature ironically leads to certain difficulties,
as long as it concerns the synchrotron 
X-ray emission of electrons. The latter, in fact,
seems to be ``over-efficient''. This, at first glance  
paradoxical statement, has a simple explanation.   
The  cooling time of electrons responsible 
for radiation of an X-ray photon  of energy $E$, 
$t_{\rm synch} \simeq 1.5 \   
B_{\rm mG}^{-3/2}  (E/1 \ \rm keV)^{-1/2} \ \rm yr$   
is small  (for any reasonable magnetic field 
$B \geq 0.01$ mG),  even  compared with the minimum 
available  time  -- the light travel  time across  the source,
$R/c \sim  3 \times 10^3  \ \rm yr$.   
This simple estimate has the following 
interesting implications. 

\vspace{2mm}
(a)  The acceleration of electrons should 
take place throughout entire volume of the 
hot spot,  otherwise the short  
propagation lengths of electrons  would not 
allow   formation of diffuse X-ray emitting regions  
on kpc scales as it is observed by Chandra. 
However,  operation of  huge quasi homogeneous 3-dimensional 
accelerators of electrons  with linear size of 1 kpc or so,   
seems to be  a serious theoretical challenge. 
Possible alternatives could be (1) 
a hypothesis of superposition of 
{\it many but compact accelerators} within the 
observed knots, or (2)
a scenario when the cloud of TeV  electrons,
accelerated in a relatively compact  region, 
expand  with the speed of light, e.g. in the form of 
a (quasi) symmetric  relativistic ``hot'' wind\footnote{The  
``hot wind''   means  that the electrons in the frame of the 
wind have a broad relativistic momentum distribution 
which would  enable  their synchrotron radiation.}.   

\vspace{1mm}
\noindent
(b)  We should expect quick formation of the 
spectrum of cooled  electrons with power-law  index  
$\Gamma_{\rm e}=\Gamma_0+1$, 
where $\Gamma_0$ is the   
acceleration  spectrum index.  The power-law index of  
radiating electrons $\Gamma_{\rm e}$ is found from the 
spectral index of synchrotron radiation $\alpha$
($S_\nu \propto \nu^{-\alpha}$) as  
$\Gamma_{\rm e}=2 \alpha+1$,  therefore $\Gamma_0=2 \alpha$. 
For some extragalactic jets, in particular 
for the knot ``A1''  of  3C~273 with a power-law 
spectral index in the X-ray band  
$\alpha_{\rm x} \approx 0.6$ (Marshall et al. 2001),   
this would imply an extremely hard  acceleration spectrum 
with $\Gamma_0=1.2 \pm 0.1$ (!) which would   
challenge any (in particular, shock) 
acceleration scenario. Moreover, the reported 
flux upper limit at the optical band 
($\leq 0.18 \mu \rm Jy$ at  $6.8 \times 10^{14}$ Hz) 
compared with the X-ray flux 
($0.018 \mu \rm Jy$ at 2 keV) 
from the so-called  $25^{\prime \prime}$  knot of  3C 120 
(Harris et al. 1999) implies that the X-ray 
to optical spectrum is  flatter than 
$S_\nu \propto \nu^{-0.35}$, and therefore 
the spectrum of radiating electrons 
is  flatter than $N(E_{\rm e}) \propto E_{\rm e}^{-1.7}$.
Harris (2001) has interpreted this result as a 
deviation from the canonical shock acceleration  
scenario but still in accordance  with oblique shock 
acceleration models 
(e.g.  Kirk 1997). However, 
the $E_{\rm e}^{-2}$  type differential electron 
spectrum is, in fact, the flattest possible  
(independent of the acceleration model) cooled 
steady-state spectrum of electrons,  as long as the 
synchrotron (and/or inverse Compton) energy losses 
play dominant role  in its  formation. Therefore, in
order to explain  the observed spectrum of synchrotron 
radiation  with  $\alpha_{\rm o-x} < 0.5$ we must 
keep the acceleration spectrum $Q(E_{\rm e})$ unchanged,
i.e. avoid radiative losses.  
Formally this could be possible within the above 
mentioned  hypothesis of fast, with {\it speed-of-light 
escape} of electrons in a form  of above mentioned 
``hot''  relativistic  wind. This is, however,
an {\it ad hoc} assumption which requires thorough 
theoretical and observational  inspections.  
 
In this paper we suggest  a different  approach 
for interpretation of the diffuse nonthermal emission 
from large scale extragalactic jets at X-rays     
and, perhaps, also at lower frequencies.  Namely, we 
assume that this radiation has a synchrotron origin, 
but is produced by very high energy  {\em protons}.
We will show that adopting magnetic fields in jet 
structures somewhat larger  (by a factor of 10 or so)   
than the field assumed  in the standard electron 
synchrotron or inverse Compton models, and speculating 
that the protons are accelerated in the jet structures 
to energies $E_{\rm p} \sim 10^{18} \ \rm eV$ or more,  
we can construct  a model which  allows an efficient 
cooling of protons via  synchrotron  radiation on very 
``comfortable''  timescales,  namely on timescales  
comparable with the (diffusive) escape time of protons 
and/or  the age of the jet. This provides effective 
propagation of protons over the entire jet structures 
on  kpc scales, and thus can naturally  explain the 
diffuse character of  X-ray emission, as well as the 
broad range of spectral indices of X-rays observed from 
different objects.  Yet, as long as the {\em proton 
synchrotron cooling time},  
the {\em escape time of particles} and the 
{\em age of the jet} are of the same order of 
magnitude, the proton synchrotron model offers a  
quite high (from 10 to  almost 100 per cent)  
efficiency of transformation of the  kinetic 
energy  of protons to  hard synchrotron X-ray emission. 
This makes, in our view, the proton synchrotron 
radiation an attractive and viable mechanism for 
interpretation of nonthermal 
X-ray emission from  large-scale extragalactic jets. 
     
\section{Extremely high energy protons in the AGN jets}
The jets of powerful radiogalaxies and AGN 
are one of a few potential sites in the Universe 
where protons could be accelerated to highest 
observed energies  of about  $10^{20} \ \rm eV$
(see e.g. Hillas 1984;  Cesarsky 1992; 
Rachen \& Biermann 1993; Henri et al. 1999). 
Dissipation of bulk kinetic energy and/or  the Poynting 
flux of jets results in, most likely through strong terminal 
shocks, generation/amplification of magnetic fields,   
heating of the ambient plasma, and acceleration of 
particles in the knots and hot spots. More  specifically,
Rachen and Biermann (1993) have demonstrated
that mildly-relativistic  jet  terminal shocks in 
hot spots of FRII radiogalaxies with typical magnetic 
field of about 0.5 mG and size $\sim 1$ kpc are indeed 
able to accelerate protons up to  $10^{20} \ \rm eV$. 
Ostrowski (1998) has shown that very effective 
acceleration of extremely high energy protons (with a  
rate of $\sim 10$  per cent of the maximum possible 
acceleration rate, $t_{\rm acc}^{-1}=c/r_{\rm g}$)  
can  take place also at the jet shear boundary layers. 

\subsection{Secondary electrons}
The undisputed synchrotron origin of nonthermal radiation 
of many knots and hot spots extending  to (at least)  
optical wavelengths is treated as a clear  evidence 
for acceleration of electrons to TeV energies.   
Besides, relativistic electrons are 
produced also in interactions of high energy 
protons with ambient gas and photon fields. 
Biermann and Strittmatter (1987) has proposed that 
the secondary component of electrons produced  by extremely 
high energy protons interacting with local photon fields, 
may play an important  role in formation of the
nonthermal radiation of radio jets. 
This idea has been further developed within the 
PIC (Proton Induced Cascade) model  by Mannheim et al. (1991) 
for interpretation of  
radiation of some prominent  radio jets features like  
the  west hot spot in Pictor A and the knot A1 in 3C 273. 
The intrinsic feature of this model is that it produces 
a distinct maximum in the spectral energy distribution 
(SED) $\nu S_\nu$ at MeV energies, and a standard hard 
X-ray spectrum  with a spectral index 
$\alpha_{\rm x} \sim 0.5$ at keV energies. 
The recent Chandra  observations of Pictor A 
(Wilson et al. 2001) give a flux which at 
1~keV appears almost 3 orders 
of magnitude below the PIC-model 
predictions.  The predictions of 
this model for  the knot A1  of  3C 273 also fall well 
below the reported Chandra flux at 1 keV 
(Sambruna et al. 2001; Marshall et al. 2001).  
Although it is formally possible to increase 
the X-ray fluxes, assuming significantly 
larger  power in accelerated protons or speculating 
with denser target photon fields, this 
would however face serious problems with the 
required nonthermal energy budget. Moreover, 
the flat X-ray spectrum of the hot spot of 
Pictor A with a spectral index $\alpha = 1.07 \pm 0.11$ 
(Wilson et al. 2001) excludes the PIC model which predicts 
much harder X-ray  spectrum. 
The X-ray spectral index  of the 
knot A1 in 3C 273 $\alpha_{\rm x} = 0.60 \pm 0.05$ is 
closer to the PIC model predictions, but in this source the 
efficiency of the process is  too low to  
explain the  detected absolute X-ray flux.  

For a broad, e.g. power-law spectrum of target photons, 
the `photo-meson' cooling time of protons can be estimated as  
$t_{\rm p \gamma}=[c <\sigma f> n(\epsilon^\ast) \epsilon^\ast]^{-1}$,
where $<\sigma f> \simeq 10^{-28} \ \rm cm^2$ 
is the photo-meson production  cross-section weighted 
by inelasticity at the photon energy 
$\sim 300 \ \rm GeV$ (see e.g. M\"ucke et al. 1999) 
in the proton rest frame and 
$\epsilon^\ast=0.03 E_{19}^{-1} \ \rm eV$;
$E_{19}=E_{\rm p}/10^{19} \ \rm eV$ is the energy 
of the proton in units of $10^{19} \ \rm eV$. The fluxes of 
jet features at most relevant for photo-meson production 
frequency band $10^{12}$-$10^{14} \ \rm Hz$  
typically are poorly known. Therefore it is 
convenient to present the low-frequency photon 
flux density in the following 
form  $S_\nu=S_0  \nu_{\rm GHz}^{-\alpha}$, 
where  $S_0$ is the flux density at 1 GHz in units of Jy. 
This  gives  
\begin{equation}
t_{\rm p \gamma} \simeq 
10^{9} c_\alpha S_0^{-1} E_{19}^{-\alpha} \theta^2 \ \rm yr ,
\end{equation}  
where  $c_\alpha \approx 0.42$;~ 3.9~ and 36 for
$\alpha=0.5$, 0.75 and 1.0, respectively;  
$\theta=R/d$ is the angular radius of the source 
in arcseconds. For example, in the A1 knot of 3C~273 with  
$\theta \sim 0.5 - 1$ arcsecond, $S_0 \simeq 0.05$Jy 
and  $\alpha \simeq 0.75$ (Marshall et al. 2000), 
the characteristic photo-meson 
production time appears more than $2 \times 10^{9} \ \rm yr$
even for $10^{20} \ \rm eV$ protons. This is too large 
compared with the escape time of highest energy 
protons from the X-ray production region. Indeed, 
the particle escape cannot be longer than 
the time determined by diffusion in the Bohm limit,
$t_{\rm esc} \approx R^{2}/2 D$ with diffusion coefficient 
$D(E)=\eta r_{\rm g} c/3$, where 
$r_{\rm g}=E_{\rm p}/eB$ is the 
the gyroradius, and $\eta \geq 1$ is the so-called 
gyro-factor (in the Bohm limit $\eta=1$). Thus
\begin{equation}
t_{\rm esc} \simeq 4.2 \times 10^{5} \eta^{-1} B_{\rm mG} 
R_{\rm kpc}^2  E_{19}^{-1} \rm \ yr. 
\end{equation}
Thus,  representing the source 
radius in the form 
$R=\theta \cdot d \simeq 4.8 \theta (d/1 \ \rm Gpc) \ \rm kpc$,
for 3C 273 ($z=0.158$) we find\footnote{Hereafter 
for the Hubble constant we adopt $H_0=60 \ \rm km/s \ Mpc$.} 
\begin{equation}
t_{\rm esc}/t_{\rm p \gamma} \simeq 8 
\times 10^{-5} \eta B_{\rm mG} E_{19}^{-0.25}.
\end{equation}
This ratio implies a very limited efficiency of 
transformation of the kinetic energy of 
protons to nonthermal radiation. Note that both 
characteristic times,  
$t_{\rm esc}$  and $t_{\rm p \gamma}$ are 
proportional to  $R^2$, consequently  the ratio
$t_{\rm esc} / t_{\rm p \gamma}$
does not depend on $\theta$, thus we cannot 
increase the photo-meson production efficiency 
assuming that the acceleration and radiation take  
place in smaller (but numerous) regions inside
the knot. The ratio slightly depends also on the 
energy of protons. Although the photo-meson production 
time decreases with energy, 
the escape time decreases even faster, thus
the efficiency  of the 
process ($t_{\rm esc}/t_{\rm p \gamma}$) becomes, 
in fact, less at higher energies.
The direct  (Bethe-Heitler)  production of $(e^+,e^-)$ 
pairs cannot significantly enhance the efficiency of 
$p \gamma$  interactions; although this process 
requires lower energy protons which are 
confined more effectively, for a broad-band 
target  radiation this process typically is  
slower than the photo-meson production.
The fluxes of {\em mm} radiation in some  hot spots, 
e.g. in the radiogalaxies Pictor A (Wilson et al. 2001) and 
Cygnus A  (Wilson et al. 2001) are  2 or 3 orders 
of magnitude higher than in the knot A1 of 3C 273,  
therefore the efficiency of $p \gamma$ interactions 
in these objects could approach to a few per cent,  
provided that particle escape takes place in 
the Bohm regime. Even so, the PIC remains a rather 
ineffective  scenario for production of X-rays,  
because it allows less than 1 per cent of the overall 
luminosity of nonthermal radiation  to be  released  
at keV energies.         

The low-frequency radiation converts,
through photon-photon interactions, 
a fraction of high energy $\gamma$-rays, 
before they escape the knots, into $(e^+,e^-)$ 
pairs. Thus, these interactions could initiate, 
in principle,  electromagnetic cascades in the jet. 
However, it is  easy to show that for most of 
the X-ray features detected  by  Chandra, this 
process could be neglected. Indeed, for the power-law 
spectrum of low-frequency radiation with above 
defined normalization, the optical depth which 
characterizes the efficiency of  the cascade 
development at $\gamma$-ray energy $E$ is estimated as   
\begin{equation}
\tau_{\gamma \gamma} \approx  \tau_{\alpha} S_0 
\theta^{-2} R_{\rm kpc} (E/{\rm 1 \ TeV})^{\alpha} \ , 
\end{equation}
with $\tau_{\alpha} \simeq 4 \times 10^{-3}; 
1.8 \times 10^{-4}$ and $8.2 \times 10^{-6}$
for $\alpha=$0.5, 0.75 and  1.0, respectively.
For example,  in  the knot A of 3C 273 with 
$\theta \sim 0.5$ arcsec, $\alpha \simeq 0.75$ 
and $S_0 \simeq 0.05$~Jy, the optical depth 
exceeds 1 only at $E \geq 10^{18}$ eV.
Thus, the cascade could be  
initiated  only at the presence of very high energy 
$\gamma$-rays. Even so, for the typical 
magnetic fields  $B \geq 0.1$ mG in the 
knots  and hot spots,  the Compton cooling of the secondary 
electrons is a much slower process  
compared with  the synchrotron losses 
($w_{\rm B} \gg w_{\rm r}$), thus in these objects the 
cascade development is strongly suppressed.   
      
Secondary electrons can be  produced also at 
interactions of relativistic protons with the 
ambient gas.  The characteristic 
cooling time of this process in the 
medium with gas number density $n_0$ 
is almost independent of the proton energy 
\begin{equation}
t_{\rm pp} =(n_0 \sigma_{\rm pp} f  c)^{-1}
\simeq 1.7 \times 10^{8} n^{-1}  \rm \ yr \ . 
\end{equation}
Because the X-ray synchrotron radiation in the 
jet could be produced by $\leq 10 \, \rm TeV$
electrons, the $pp$ interactions 
require relatively low energy protons,   
$E_{\rm p} \leq 100 \  \rm TeV$ the escape time
of which may exceed the jet age.  
On the other hand  the $pp$  cooling time  of protons 
is quite large, therefore this attractive 
mechanism, which produces relativistic electrons
throughout  the entire jet structure,  may have an impact on 
the overall nonthermal radiation, provided that  
the ambient gas density in the knots and hot 
spots exceeds  $0.1-1 \ \rm cm^{-3}$.    
 
\subsection{Synchrotron radiation of protons}
During  the jet lifetime most of the energetic particles  
escape the knots. In particular,  in the Bohm regime, 
when the particles tend to drift away 
most slowly ($\eta=1$), the break in the initial 
proton spectrum takes place at 
$E_{\rm p} \simeq 4.2 \times 10^{17} 
B_{\rm mG} R_{\rm kpc}^2 
(\Delta t/10^7 \ \rm yr)^{-1} 
  \rm \ eV$. Since the observations limit 
the size of a typical  knot or a  hot spots 
by $R \leq$ few kpc, the 
only possibility to move the break point to higher energies is
the increase of the magnetic field, $B \geq 1 \ \rm mG$.
However, at such large magnetic fields the 
proton synchrotron radiation (see e.g. Aharonian 2000) 
becomes an additional, or even more important limiting 
factor with characteristic cooling time
\begin{equation}
t_{\rm synhc} \simeq 1.4 \times 
10^{7} B_{\rm mG}^{-2} E_{19}^{-1} \ \rm yr\ ,
\end{equation}
Remarkably, the radiation by highest energy protons 
is  released in the X-ray domain:
\begin{equation}
h \nu_{\rm m}=0.29 h \nu_{\rm c} \simeq 2.5 \times 10^{2} 
B_{\rm mG} E_{19}^{2} \ \rm keV .
\end{equation}       
Correspondingly, the characteristic time of radiation of a 
proton-synchrotron photon of energy $E$ is 
\begin{equation}
t(E) \simeq 2.2 \times 10^{8} 
B_{\rm mG}^{-3/2} (E/1 \ \rm keV)^{-1/2} \ \rm yr \ .
\end{equation} 
Taking into account that 1 keV  photons are produced 
by protons with energy 
$E_{19} \simeq 0.063 \ B_{\rm mG}^{-1/2}$,
the time of radiation of synchrotron X-ray photons becomes 
less than the time of particle escape in the Bohm regime
if $B_{\rm mG}^3 R_{\rm kpc}^2 \geq 30$. 
For a typical size of jet knots of about 1 kpc, 
this would require $B \geq 3 \times 10^{-3} \ \rm G$ 
and correspondingly the total energy contained in the 
the magnetic field  
$W_{\rm B}=B^2 R^3/6 \geq  4 \times 10^{58} \ \rm erg$. 
Interestingly, in such an extreme regime, when almost the 
whole kinetic energy of  accelerated protons is released 
in synchrotron X-rays for a typical time of about 
$\leq  4 \times 10^7$ yr,  the luminosity  
of synchrotron radiation  at 1 keV  is expected at the level 
$L_{\rm X} \simeq 3 \times 10^{43} \kappa \ \rm erg/s$,
assuming equipartition between the magnetic
field and accelerated protons, $W_{\rm p}=W_{\rm B}$.
The parameter  $\kappa$ is the fraction of the kinetic  
energy in protons responsible  for synchrotron X-rays, 
$E_{\rm p} \geq 10^{17} \ \rm eV$, to the total energy. 
For a $E_{\rm p}^{-2}$ type proton spectrum extending beyond 
$10^{18} \, \rm eV$,  $\kappa \sim 0.1$.
Note that in the regime dominated by proton synchrotron 
losses, the total luminosity of the proton-synchrotron 
radiation does not depend on the magnetic field, and 
is determined simply by the acceleration power of 
highest energy protons, 
$L_{\rm x} \simeq L_{\rm p}(\geq E_{\rm p}^\ast)$. 

Below we present detailed numerical calculations for several 
prominent  extragalactic jet features.  
We assume that during the lifetime of the jet $\Delta t$, 
protons are injected into a homogeneous and spherically 
symmetric region, with acceleration spectrum 
$Q_{\rm p}(E_{\rm p})=Q_0 E_{\rm p}^{-\Gamma_{\rm p}} 
\exp{(-E_{\rm p}/E_0)}$ 
and (quasi) continuous constant rate $L_{\rm p}=
\int Q_{\rm p}(E_{\rm p})E_{\rm p} {\rm d}E_{\rm p}$.   
The kinetic equation for energy distributions of 
relativistic particles which takes into account 
the radiative and escape losses,  
and the time-dependent solutions to this equation, are
described in the papers by 
Atoyan \& Aharonian (1999) 
and Aharonian \& Atoyan (1999). 
In this study the interactions of protons 
with the magnetic field 
({\it synchrotron radiation}), photon fields 
({\it photo-meson- and  Bethe-Heitler pair production}) 
and the  ambient gas 
($pp \rightarrow \pi \rightarrow e^+e^-\gamma$), 
as well as the {\it synchrotron radiation} and 
{\it inverse Compton scattering} of 
secondary  electrons are included in calculations. 
The fast synchrotron cooling of secondary electrons
in relatively strong magnetic fields allow us to 
ignore  the cascade processes (initiated by 
$\gamma$-$\gamma$ interactions) without any significant 
impact on accuracy of calculations. 

\section{Jet features in Pictor A, PKS 0637-752, and 3C 120}
The luminosities of proton-synchrotron radiation
calculated  for 3 prominent jet features --
the western hot spot in Pictor A, 
the radio knot at $25^{\prime \prime}$ 
distance from the nucleus of the Seyfert galaxy 3C 120, 
and  the bright X-ray region in the inner western jet of 
PKS 0637-752 (at $\sim 8^{\prime \prime}$ from the core) -
are shown in Fig.~1.   
The shape  of the resulting spectrum of protons 
primarily depends on the index of acceleration spectrum 
$\Gamma_0$, the magnetic field $B$ and the size $R$.
The total power in the accelerated protons $L_{\rm p}$
determines the absolute X-ray flux. For clearness, 
in Fig.~1 we present the results of calculations which 
take into account interactions of protons with 
universal 2.7 K CMBR, but ignore the interactions 
with the ambient gas and local photon fields. 

\subsection{Pictor A} 

The western X-ray hot spot at  $4^{\prime}.2$ from the 
nucleus of this nearby ($z=0.035$) powerful radiogalaxy 
coincides with  the radio jet and is laterally extended by 
$\sim 2 \ \rm kpc$ (Wilson et al. 2001).  The X-ray 
emission  of the hot spot is well described  by a 
flat power-law spectrum with a photon index  
$\alpha+1=2.07 \pm 0.11$.   
This implies that within the proton-synchrotron  model the 
spectral index  of  radiating  protons should be  close to  3. 
Thus,  for the shock  acceleration spectrum with  canonical  
power-law  index $\Gamma_0 \sim 2$ we must allow  
an increase  of the spectral  slope by 1.  Such a steepening 
can be naturally provided by  an  intensive synchrotron  
cooling and/or by fast energy-dependent escape of protons 
from the source with $\tau_{\rm esc}(E)  \propto 1/E$. 
In the synchrotron-loss 
dominated regime the corresponding break in the 
proton-synchrotron spectrum  (the point where  
$S_\nu \propto \nu^{-0.5}$ 
is changed to $\propto \nu^{-1}$) takes place at
\begin{equation}
(h \nu)_b \simeq 5 \ B_{\rm mG}^{-3} 
\Delta t_8^{-2} \ \rm keV \ ,
\end{equation}
where $\Delta t_8=\Delta t/10^8 \ \rm yr$ is the 
duration of the particle acceleration in units of 
$10^8 \ \rm yr$. For a typical lifetime of the jet 
(the period of activity of the central engine) 
$\Delta t \leq 10^8  \, \rm yr$, the flat
($\nu S_\nu=const$)  spectrum of the hot spot  
above 1 keV would require  magnetic field larger than 1 mG. 
In the regime dominated by particle escape, 
the break in the spectrum depends on the propagation character 
of protons. In particular in the regime of Bohm diffusion 
the break in the proton-synchrotron  spectrum appears at 
\begin{equation}
(h \nu)_b \simeq 0.004 \ B_{\rm mG}^{3} 
\Delta t_8^{-2}  R_{\rm kpc}^{4} \eta^{-2} \ \rm keV \ .
\end{equation} 

Note that the characteristic times of synchrotron cooling and 
the particle escape in the Bohm regime have 
the same $\propto 1/E$ dependence, therefore the 
dominance of  synchrotron losses 
does not depend on the specific energy interval, 
and is  determined by  
the following simple  condition:
\begin{equation}
B_{\rm mB}^3 R_{\rm kpc}^{2} \eta^{-1} \geq 30 \ .
\end{equation} 

In Fig.~1 we show two ``proton-synchrotron'' fits to 
the spectrum of the hot spot of Pictor A calculated  
({\it a}) for a regime dominated by {\it synchrotron losses}  
with two key model parameters $B =$3 mG and $R=2 \ \rm kpc$ 
(curve 1),  and ({\it b}) for a regime  dominated by 
{\it particle escape} with $B =$ 1 mG and  $R=1 \ \rm kpc$ 
(curve 2).
In both  cases the following  additional  model 
parameters were  assumed:
gyro-factor  $\eta=1$  (Bohm diffusion);  power-law index 
$\Gamma_0=2$ and 
maximum energy of accelerated protons 
$E_0=10^{20} \ \rm eV$.
And finally, to match the absolute  
X-ray fluxes we assume that the proton 
acceleration  takes  place in a (quasi) continuous regime
during last  $\Delta t =10^{8} \ \rm yrs$ with  
rates  $L_{\rm p}=4.5 \times 10^{43}$ 
and $1.35 \times 10^{45} \rm erg/s$ for the cases 
({\it a}) and ({\it b}),  respectively. 
%
\begin{figure}
\epsfxsize=8.0 cm
\centerline{\epsffile[106 30 475 350]{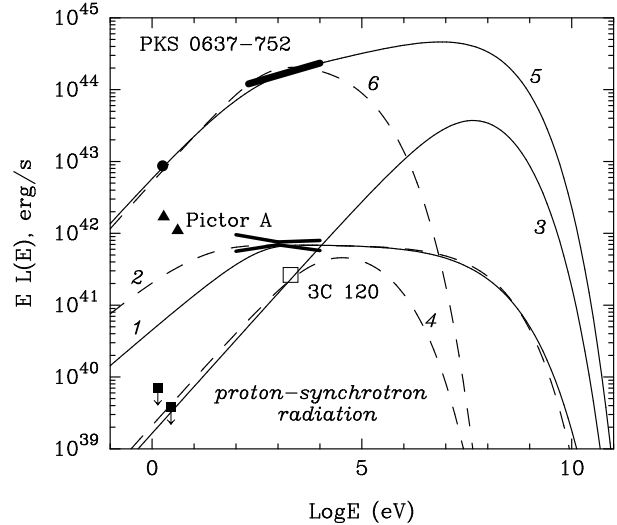}}
\caption{Proton synchrotron luminosities of jet features in 
Pictor A (western radio hot spot), 3C 120 
($25^{\prime \prime}$
radio knot), and PKS 0637-752 (the outer jet component). 
The model parameters are  described in the text. 
The optical and X-ray  data are from 
Wilson et al. (2001) for Pictor A; 
Harris et al. (1999) for 3C 120; 
Schwartz et al. (2000) for PKS 0637-752.
}
\label{fig1}
\end{figure}                                               

Although both spectra  satisfactorily fit the reported 
X-ray data above 0.5 keV\footnote{Below 0.5 keV the spectrum 
corresponding to the case ({\em a}) passes below the range of 
extrapolated fluxes given by Wilson et al. (2001).  
However  better agreement at these  energies could be  
easily achieved  assuming a slightly 
larger magnetic field.},  the case ({\it a}) is 
more attractive on energetic grounds --   
it requires $\sim 30$ times  less power in accelerated 
protons.  Note that in the regime dominated by synchrotron 
losses, the efficiency of X-ray production is close to 100 
per cent,  the  luminosity  being almost independent of 
the magnetic field, $L_{\rm x-\gamma} \sim L_{\rm p}$.  
In the particle-escape dominated regime, 
$L_{\rm x-\gamma} \propto B^{2}$.

The  spectra of relativistic protons 
$W_{\rm p}(E_{\rm p})=E_{\rm p}^2 N(E_{\rm p})$
trapped  in the hot spot are shown in Fig.~2.  Like 
Fig.~1,  the curves 1 and 2  correspond
to  ({\it a}) and  ({\it b}) cases, respectively. 
The spectra are similar with an original (acceleration)  
shape $\propto E_{\rm p}^{-2}$ below the break energy
at which  the escape and synchrotron cooling times
exceed  the age of the source, and $\propto E_{\rm p}^{-3}$
 at higher energies. At the same time, while  in the case 
({\it a}) the total energy in protons is rather modest, 
$W_{\rm p}^{(\rm tot)}=\int W(E_{\rm p}) 
{\rm d} E_{\rm p} \simeq 1.1 \times 10^{59} \ \rm erg$, 
in the case ({\it b}) it is significantly higher, 
$W_{\rm p}^{(\rm tot)} \simeq 2.9 \times 10^{60} \ \rm erg$.
Thus, while in the case ({\it a}) the conditions 
are close to the  equipartition between the protons 
and magnetic filed 
($w_{\rm p} \simeq 1.3 \times 10^{-7} \ \rm erg/cm^3$, 
$w_{\rm B} \simeq 3.6 \times 10^{-7} \ \rm erg/cm^3s$), 
in the case ({\it b}) the pressure of protons exceeds 
the B-field pressure by 3 orders of magnitude. 

%
\begin{figure}
\epsfxsize=8.0 cm
\centerline{\epsffile[50 25 508 331]{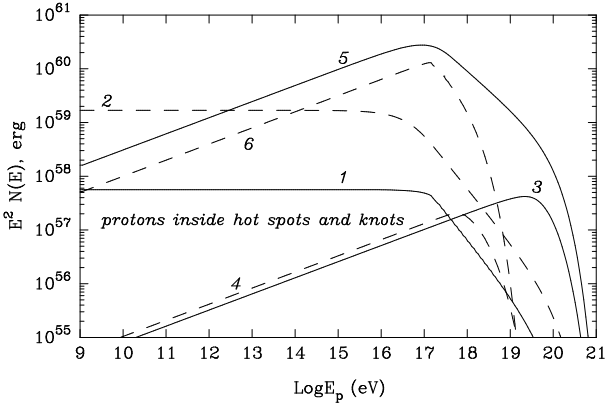}}
\caption{Total energy of relativistic protons confined  in  
X-ray emitting regions of radio jets:  
the western hot spot in Pictor A (curves 1 and 2),
$25^{\prime \prime}$ knot of 3C 120 (curves 3 and 4), and 
the outer jet component in  PKS 0637-752 (curves 5 and 6) .
}
\end{figure}

Finally we note that the proton synchrotron radiation cannot 
be responsible for the observed radio and optical fluxes.
The two optical/UV points (triangles)  in Fig.~ 1 
obtained from the {\it HST} observations of the hot spot 
(Wilson et al. 2001)  confirm the break in the spectrum 
at $\nu \sim 10^{14} \ \rm Hz$ reported earlier  by 
Meisenheimer et al. (1997). Both the slope of the  spectrum 
connecting the radio and optical fluxes,  
as well as the spectral break  above  $10^{14} \ \rm Hz$  
can be  naturally  explained  by  synchrotron 
radiation of electrons (Meisenheimer et al. 1997). 
We note, however,  that this  radiation component  
is produced,  most probably, in region(s)  separated from  
the X-ray production region,  both in {\em space} and 
{\em time}.  Indeed, the break in the spectrum of 
electron synchrotron radiation in the regime when the  latter
dominates  in the formation of the 
spectrum of parent electrons,  appears at 
$\nu_b \simeq 60  \ B_{\rm mG}^{-3} \Delta t_8^{-2} \ \rm Hz$.
Thus, the  {\em synchrotron-loss} origin of the break
at $10^{14} \ \rm Hz$ would require, for any reasonable 
magnetic field $B \geq 0.01$~mG, that the acceleration 
of this electron population took place  
recently, namely since last $10^{5}$ years or less.  
Such a conclusion could be avoided if we
assume that the electrons escape the source so fast, 
that the synchrotron losses become
negligible. The comparison of the particle escape time
\begin{equation}
t_{\rm esc} \simeq 3.2 \times 10^3 R_{\rm kpc} \ (c/v) \ \rm yr
\end{equation}
with the  radiation time of a synchrotron photon 
of frequency $\nu$,  
\begin{equation}
t(\nu) \simeq 75  B_{\rm mG}^{-3/2} 
(\nu /10^{14} \ \rm Hz)^{-1/2} yr \,
\end{equation}
lead to the conclusion that even 
for the escape of electrons with the speed of light,
the magnetic field should be less than 0.1 mG.

The  calculations presented in  Fig.~1 include 
interactions of protons with  2.7 K CMBR. The 
synchrotron radiation  of the secondary electrons peaks 
at GeV energies, with luminosities at 1 keV of about
$3.5 \times 10^{36} \ \rm erg/s$ for the case 
({\it a}) and by a factor of 20 larger for the case 
({\it b}) (not seen in Fig.~1). 
In the hot spot of Pictor A, the interactions of 
protons with the local (synchrotron) radiation  
at $\nu \sim 0.1-100$ GHz are more (by two orders 
of magnitude) frequent, especially if we  assume 
a small, sub-kpc  size of the hot spot. Even so, 
this process is not sufficiently effective to 
reproduce  the observed X-ray fluxes. Moreover,  
the synchrotron radiation  of secondary (from 
$p \gamma$ interactions) electrons at 1 keV has, independent 
of the proton spectrum,  standard spectrum with spectral 
index $\simeq 0.5$, i.e. significantly harder than the 
spectrum detected by Chandra (Wilson et al. 2001).  

\subsection{3C 120} 

The X-radiation  from a  radio knot  
at a distance of  $25^{\prime \prime}$ from the nucleus of
the Seyfert galaxy 3C 120  found in the ROSAT HRI data  
(Harris et al.  1999) reserves a special interest.    
The flux  predicted by the SSC models appears several 
orders of magnitude below the observed  X-ray flux. 
The two-component synchrotron model which postulates 
two different populations  of relativistic 
electrons responsible for the radio-to-optical
and X-ray emissions,  at first glance seems a 
natural interpretation. However, the reported flux  
limit at optical frequencies (Harris et al. 1999) 
challenges this model as well. Indeed, the upper limit 
at $\nu=6.8 \times 10^{14} \ \rm Hz$
($S_\nu  \leq 0.18 \mu \rm Jy$) combined with the 
X-ray flux at 2 keV   ($0.018 \mu \rm Jy$) 
(see Fig.~1) requires  an unusually hard  spectrum 
between the optical and X-ray band,
$S_\nu \propto \nu^{-\alpha}$ with  $\alpha \leq 0.35$.
This implies that the  spectrum of radiating electrons  
should be flatter than 
$N(E_{\rm e}) \propto E_{\rm e}^{-1.7}$.
Although such a spectrum deviates from predictions 
of the canonical shock acceleration theory, formally
we cannot exclude other, more effective 
particle acceleration scenarios.  
However,  as long as the synchrotron energy losses 
(${\rm d}E_{\rm e}/{\rm d}t \propto E_{\rm e}^2$) 
play dominant role in formation of the electron 
spectrum, the  latter cannot be harder, independent 
of the initial (acceleration) spectrum  $Q( E_{\rm e})$,  
than $N(E_{\rm e}) \propto E_{\rm e}^{-2}$.  
Indeed, since the cooled 
steady-state  spectrum of electrons is determined as 
\begin{equation}
N(E_{\rm e}) \propto  ({\rm d}E_{\rm e}/{\rm d}t)^{-1}   
\int _{E_{\rm e}} Q(E_{\rm e}){\rm d}E_{\rm e},
\end{equation}
even in the extreme case of abrupt low-energy cutoff 
in the injection spectrum (i.e. 
$Q(E_{\rm e})=Q_0 E_{\rm e}^{-\Gamma_0}$
at $E_{\rm e} \geq E^\ast$, and $Q(E_{\rm e})=0$ at 
$E_{\rm e} \leq E^\ast$), the spectrum of cooled electrons has
a {\em  broken power-law} shape with 
$N(E_{\rm e}) \propto E_{\rm e}^{-(\Gamma_0+1)}$ at
$E_{\rm e} \geq E^\ast$,
and  $N(E_{\rm e}) \propto E_{\rm e}^{-2}$ 
at $E_{\rm e} \leq E^\ast$.  The acceleration spectrum
$Q(E_{\rm e})$ can be sustained  unchanged, 
assuming fast, with speed of light, 
escape or adiabatic losses of electrons. 
However, even with such a 
dramatic  assumption, the electron escape time,  
$t_{\rm esc}=R/c \simeq 3  \times 10^3 R_{\rm kpc} \ \rm yr$,
in  the knot with a radius 2-3 kpc appears 
longer than the typical lifetime of electrons responsible 
for synchrotron radiation of  optical and X-ray photons,      
$t_{\rm synch} \simeq 1.5 B_{\rm mG}^{-3/2} (E/1 
\ \rm keV)^{-1/2} \ \rm yr$,
unless the magnetic field is less than 
$\sim 10^{-5} \ \rm  G$,  or we should assume 
(Harris et al. 1999) that the observed X-ray 
flux is a result of  contributions from   
short-lived (transient) compact regions
where the energy-independent  escape  
losses could dominate over the synchrotron losses.  

We do not face such a problem with the synchrotron 
radiation of protons. For a  reasonable set of  
model-parameters, the  proton escape could dominate  
over the synchrotron cooling, but yet the luminosity 
of the proton-synchrotron radiation could be 
maintained at a rather  high level. Obviously, 
quite similar to the electron synchrotron model, 
the comparison  of the X-ray and optical fluxes tells us that
the  spectrum of the radiating protons should be harder than   
$E^{-1.7}$. This implies that for even hardest 
possible acceleration spectra  $\propto E^{-1.5}$ 
(Malkov 1999), the escape of particles should be
essentially energy-independent 
($t_{\rm esc} \propto E^{-\beta}$
with $\beta \leq 0.2$).
In Fig.~1 we show a possible version of the 
proton-synchrotron radiation calculated for the 
following model parameters: the magnetic field 
$B= 1.5$ mG, radius of the X-ray production region 
$R=3$ kpc,   acceleration spectrum of protons with 
$\Gamma_0=1.7$ and  $E_0=10^{20} \ \rm eV$,
energy-independent escape of protons with  
$t_{\rm esc}=5 \times 10^{5} \ \rm yr$. For 
the diffusive propagation of particles, the latter 
corresponds to the diffusion coefficient 
which at proton energies $\geq 10^{17}$ (responsible for
production of 0.1-10 keV synchrotron photons) 
is much larger than the  Bohm diffusion coefficient.  
This provides the dominance of escape losses,
and thus does not allow deformation of the primary 
(acceleration) spectrum of protons up to highest 
energies (see curve 3 in Fig.~2) . 
Then,  the efficiency of the proton-synchrotron 
radiation, which  is characterized by the ratio 
of the escape time to the synchrotron  cooling time,  
at 1 keV is about $\approx 0.4$ per cent.  The efficiency is 
much higher (close to 100 per cent)  for  100 MeV photons 
(see curve 3 in Fig.~1) produced by  highest  energy 
($\sim 10^{20} \ \rm eV$)  protons. The corresponding
flux of 100 MeV $\gamma$-rays 
$J_\gamma \sim 5 \times 10^{-8} \ \rm ph/cm^2 s$
is sufficiently large  to be detected by GLAST 
(Gehrels \& Michelson  1999).      

To reproduce  the observed  X-ray flux at 1 keV 
we must assume a rather high injection rate of  
accelerated protons during $10^{8}$ yr operation 
of the jet,  $L_{\rm p}=1.65 \times 10^{45} \ \rm erg/s$.  
Note, however,  that  $\geq  10^{18}$~eV
protons  do not contribute, for the given magnetic 
field of about 1.5 mG, to the production of  
$\leq 1$ keV photons. Therefore we may somewhat  
(by a factor of 4) soften this  requirement 
by reducing  the  high energy cut-off in the proton 
spectrum down to $E_0=10^{18}$ eV. Further 
significant  reduction of the required acceleration 
power could be achieved assuming more effective 
particle confinement. In Fig.~ 1 we show the 
luminosity of the proton-synchrotron radiation 
(curve 4)  calculated for 
$t_{\rm esc} =5 \times 10^{7} \ \rm yr$\footnote{In 
order to avoid violation of the Bohm limit at 
higher energies, it is assumed that above $10^{18}$ eV   
the escape time decreases as $\propto 1/E$.} and  
$E_0=5 \times 10^{18}$ eV, but sustaining all other  
model parameters unchanged. In spite of the increase 
of the escape time by two orders of magnitude, 
which becomes comparable with the 
age of the source $\Delta t=10^{8}$ yr,  
the synchrotron cooling remains a slower process, 
compared with  the particle escape,   until 
$E_{\rm p}=1.2 \times  10^{18}$ eV. Therefore 
the break in the synchrotron spectrum occurs only at 
$\sim 5 \ \rm keV$.  At the same time, the required 
proton acceleration power becomes rather modest; now only 
$L_{\rm p} \simeq 1.0 \times 10^{43} \ \rm erg/s$
is needed to explain the observed X-ray  flux.  
 
\subsection{PKS 0637-752}

This quasar at a redshift $z=0.651$ has the largest 
and most powerful X-ray jet  detected so far 
(Chartas et al. 2000,  Schwartz et al. 2000).  
The X-ray luminosity of the $\geq 100$ kpc  
jet of about $L_{\rm X} \sim 4 \times 10^{44} \ \rm erg/s$
is contributed mostly by bright condensations between 
$7^{\prime \prime}.5$ and $10^{\prime \prime}$. This  
region contains three 3 knots resolved in radio and 
optical wavelengths. The enhanced X-ray  emission 
associates,  most probably, with these knots, 
although  its profile does not exactly repeat 
the radio and optical profiles of the jet.
Therefore the combined flux of these three knots 
$S_\nu = 0.57 \ \mu \rm Jy$
at an  effective frequency 
$4.3 \times 10 ^{14}$ Hz (Schwartz et al. 2000)
should be considered as an upper limit for the 
optical flux from the X-ray emitting regions. 

Because the individual X-ray knots are not clearly resolved,
here  we consider a simplified  picture, namely treat  the 
overall emission from  the  X-ray enhanced region as  
radiation  from a single source. 
The X-ray luminosity based on the 
measured flux density at 1 keV, 
$5.9 \times 10^{-14}  \ \rm erg/cm ^2 s \ keV$ and spectral  index 
$\alpha=0.85$ (Schwartz et al. 2000) is shown,  
together with the model calculations, in Fig.~1. 
Curve 5 is 
obtained for the following parameters: $B = 1.5$ mG, 
$\Gamma_0=1.75$, $E_0=10^{20}$ eV, $R=5 \ \rm kpc$. 
Also it was assumed that particles propagate in a 
``relaxed''  Bohm diffusion regime with gyro-factor 
$\eta=10$. Under these conditions the escape losses 
dominate over the synchrotron losses, and  
become important above $10^{17}$ eV. This  
result in the steepening of the protons 
spectrum,  from the initial $E_{\rm p}^{-1.75}$ 
to $E_{\rm p}^{-2.75}$. The latter continues up 
to the exponential cutoff in the  
acceleration spectrum at $E_0=10^{20}$ eV (curve 5 in Fig.~2). 
The corresponding proton-synchrotron  spectrum nicely
fits the observed X-ray spectrum and,  
at the same time,  agrees with the optical flux 
reported from that part of the jet. Note, 
however, that the optical flux shown in Fig.~1 perhaps 
should be considered rather as an upper limit. 
Correspondingly, if the intrinsic optical flux of the 
X-ray emitting region is significantly less than this 
upper limit, we must assume flatter proton acceleration 
spectrum.  It should be noticed also that the 
extrapolation of the proton synchrotron spectrum down to 
radio wavelengths appears significantly below the reported 
radio fluxes at 4.8~GHz and 8.6~GHz  
(Chartas et al. 2000), which most probably are due to
the synchrotron radiation of directly accelerated electrons.    

For the chosen set of model parameters, the absolute 
X-ray flux requires huge  acceleration power 
$L_{\rm p}=3 \times 10^{46}$ erg/s, which however 
agrees with the estimates of the jet power 
(in the form of kinetic energy of particles 
and of Poynting flux) which in most luminous 
extragalactic objects could be as large 
as $10^{47}-10^{48}$ erg/s (see e.g.  Sikora 2000,  
Ghisellini  \&  Celotti  2001).  
We may nevertheless reduce the energy  requirements 
assuming a somewhat different model parameters which 
would minimize the escape losses. 
An example of X-ray luminosity calculated for a  favorable 
set of model parameters  is presented by curve 6 in Fig.~1. 
In particular, compared with curve 5 the magnetic field 
is increased  to 3 mG, the size is reduced to 3 kpc, and it is 
assumed that  the particle diffusion  proceeds in the  Bohm 
regime ($\eta=1$). This implies  that  
$B_{\rm mB}^3 R_{\rm kpc}^2 \eta^{-1}=243$, 
i.e. the proton losses are strongly dominated  
by synchrotron cooling.  In order to reduce the 
required proton acceleration power  by another 
factor of $\sim 3$, an ``early'' exponential 
cutoff in the acceleration spectrum is assumed 
at $E_0= 2 \times 10^{18}$ eV. The resulting proton 
spectrum shown in  Fig.~2 (curve 6) could 
be treated  as  an optimum  spectrum  designed 
to maximize the production of X-ray at 1 keV,  
as it is seen in Fig.~1. 
The required proton  power  now is 10 times 
less than in the previous case,  
$L_{\rm p} =2.9 \times 10^{45}$ erg/s.  
Further significant reduction of $L_{\rm p} $ is almost 
impossible given the fact that the X-ray luminosity in the  
interval 0.2-10 keV already is huge,
$L_{\rm x} \simeq 5 \times 10^{44}$ erg/s, unless we assume 
that the jet is moving with  a  bulk 
Lorentz factor $\Gamma_{\rm j} \gg 1$ towards the observer. 

Scarpa and Urry (1999) argued that the relativistic bulk 
motion of the jets on kpc scales, and the consequent beaming 
effect (if the jets are aligned with the line of sight) 
would allow reduction of the magnetic field, 
and consequently - longer electron lifetimes. 
This would help, at least partly, 
to remove some of the problems of electron synchrotron 
models suggested for radio and  optical jets. Actually, 
the hypothesis of relativistic beaming with Doppler factor 
$D_{\rm j} \gg 1$ implies dramatic reduction of the jet's 
intrinsic luminosity, and thus offers  significantly 
relaxed conditions to all models of nonthermal emission 
of large scale extragalactic jets.  In particular, 
Tavecchio et al. (2000) and  Celotti et al. (2001)  
have shown  that the inverse Compton  scattering of 
relatively low energy electrons on CMBR is a viable 
mechanism for explanation of X-ray emission of the jet 
of PKS 0637-752. Obviously, the beaming effect would be
a great help also for the proton synchrotron model in the 
sense of significant reduction of the required proton  
acceleration rate and/or the strength of the magnetic 
field. 

\section{Jet in   3C 273}

At least 4  features have been revealed in the 
Chandra images of the X-ray jet of  quasar 3C 273
(Marshall et al. 2000; Sambruna et al. 2001).  
Approximately $40 \%$ of the total X-ray luminosity 
of the jet is contributed by the  brightest knot 
called A1 (Bahcall et al. 1995). 
This knot is  unique also in the sense  that its  radio,  
optical and  X-ray fluxes are  consistent with  a single 
power-law spectrum (R\"osner et al.  2000, Marshall et al.  2001; 
see, however, Sambruna et al. 2001). 
The  spectral energy distribution  (SED) of the knot A1
is shown in Fig.~3. The observed fluxes  
at radio (Merlin array),  optical ({\em HST}), and X-ray 
(Chandra) bands are  from  Marshall et all (2001).
The overall slope with spectral  
index  $\alpha_{\rm r-x} \simeq 0.75$  agrees well with 
the local slope  at optical  wavelengths,  and 
matches the   X-ray flux at 1 keV, although the 
local slope  at  X-rays is slightly  different,   
$\alpha_{\rm x}  = 0.60 \pm 0.05$ 
(Marshall et all 2001).  
R\"osner et al. (2000) has interpreted 
the power-law  behavior of the 
broad-band spectrum of A1 
as a signature of synchrotron radiation 
by a single population of electrons. The spectral 
index $0.75$ requires power-law index of  
electrons $\Gamma_{\rm e}=2.5$ and therefore, 
since the synchrotron cooling time of electrons 
is short compared  with the  lifetime of the jet,  
we should assume a  hard,  
$Q(E_{\rm e}) \propto E_{\rm e}^{-1.5}$ 
type electron acceleration spectrum. 
%
\begin{figure}
\epsfxsize=8.0 cm
\centerline{\epsffile[75 427 500 727]{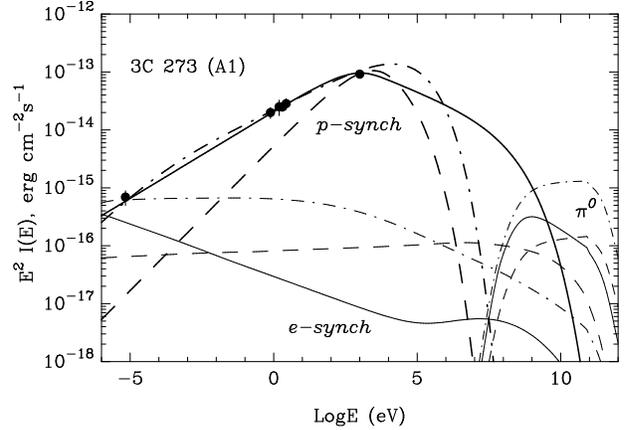}}
\caption{The spectral energy distribution of the 
knot A1 in  3C 273.  The radio, optical and X-ray 
fluxes are from Marshall et al. (2001). The solid, 
dashed and dot-dashed curves correspond to
3 different sets of model parameters discussed in the text.
The {\em heavy, standard and thin}  lines  
represent the fluxes of (1) synchrotron radiation 
of protons,  (2) synchrotron radiation of secondary 
electrons produced in  $p \gamma$ and $pp$ interactions, 
and (3) $\pi^0$-decay $\gamma$-rays produced at $pp$ 
interactions,  respectively. 
}
\end{figure}
Because of severe synchrotron losses,
the X-ray emitting  electrons could 
not propagate  far from  their birthplace/accelerator.  
To some extent  this is true also for  the electrons 
responsible for optical radiation. Therefore we should expect
a point-source type morphology, unless the electron acceleration  
takes place throughout the entire knot, 
or the radiation is contributed 
by many compact regions inside the knot.   

\subsection{Proton-synchrotron model}

The proton synchrotron model offers 
more freedom for the  principal model 
parameters. Indeed,  for the size of the knot
of about 1 kpc  and for the ambient 
magnetic field $B \geq $ 1 mG, the synchrotron-cooling 
and the escape  times  of protons are comparable with 
the age of the jet.  Thus,  with certain  assumptions  
about the acceleration spectrum and the propagation  
of protons  we can satisfactorily fit the fluxes  
as well as  to explain the kpc size of the  
observed diffuse nonthermal emission.

In Fig.~3 we show three  spectra of proton-synchrotron radiation 
modeled for the knot A1. For all 3 curves we assume that 
continuous injection of relativistic protons proceeds with a
constant rate  during last $3 \times 10^7$ yr. 
The solid curve  is calculated  for $B=5$  mG and  
$R=1$ kpc,  and assuming that the protons  
propagate in the Bohm diffusion regime 
with $\eta=1$ (model I). 
This implies that the synchrotron losses dominate 
over the particle escape, but  yet,  for the assumed source  
age and the magnetic field, the synchrotron losses become 
important only for protons responsible for X-ray flux 
above 1 keV  ($E_{\rm  p} \geq 3 \times 10^{17} \ \rm eV$).
Since the escape losses are also  small and do not 
change the  acceleration spectrum of protons either, 
in order to explain  the entire range of nonthermal emission
from radio to X-rays with spectral index 
$\alpha_{\rm r-x} \simeq 0.7$, we must assume 
a steep proton acceleration spectrum with 
$\Gamma_0=2 \alpha_{\rm r-x}+1=2.4$. The solid 
curve in Fig.~3 corresponds to such a power-law 
index of accelerated protons. The disadvantage 
of this fit  is that it requires uncomfortably 
large acceleration rate 
$L_{\rm p}=1.2 \times 10^{47} \ \rm erg/s$
with the total amount of protons deposited 
in the knot  during  $3 \times 10^7$ yrs,  
$W_{\rm p}^{\rm (tot)}  \simeq   10^{62} \ \rm erg$ 
(see Fig.~4). This implies that the energy density 
of cosmic rays  exceeds the density of the magnetic 
field by 3 orders of magnitude.

It should be noticed, however, that this  huge 
total energy is contributed by low energy
protons  which, from the point of view of 
production of radiation above $\geq$ 1 GHz,    
are in fact ``vain''  particles. Therefore assuming 
a low-energy cutoff in the acceleration spectrum, 
e.g. at $10^{13}$ eV, we can  significantly reduce 
the energy  requirements. 
Another,  perhaps a more natural way  to avoid the 
low-energy protons can be achieved assuming 
a hard  acceleration spectrum coupled with 
energy-dependent  escape of particles.  The dot-dashed 
curve  in   Fig.~4 corresponds to the  case 
of  the canonical  shock  acceleration spectrum  
with $\Gamma_0=2$,  and the 
escape time  taken  in the form 
$t_{\rm esc}(E)=1.4 \times 10^{7} (E/10^{14} 
\ \rm eV)^{-1/2}$ yr (model II). 
The resulting proton spectrum  inside the knot in 
the most relevant energy region above $10^{13} \ \rm eV$ 
becomes steeper, $E^{-2.5}$.
The assumed large magnetic field $B=10$ mG,
as well as the  upper energy cutoff at  
$E=10^{18} \ \rm eV$ in the acceleration spectrum 
allow additional reduction of the  acceleration power  
down to $1.05 \times 10^{46} \ \rm erg/s$.  
The corresponding  energy of protons confined  
in the knot now is $4.9 \times 10^{60} \ \rm erg$.   
The spectrum of proton synchrotron radiation 
calculated for this set of parameters is shown in Fig. 3
by the dot-dashed curve.  
%
\begin{figure}
\epsfxsize=8.0 cm
\centerline{\epsffile[50 405 480 708]{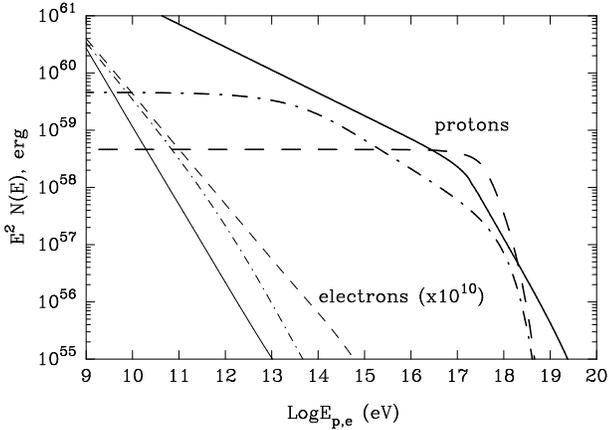}}
\caption{Total energy of relativistic protons trapped in  
the knot A1 of the quasar 3C 273.  The heavy  
solid, dashed, and 
dot-dashed curves correspond to 3 different combinations 
of the model parameters as in Fig.~3. The fluxes of secondary 
$p \gamma$ and $pp$ electrons multiplied by $10^{10}$ 
are also shown (thin lines).  
}
\end{figure}

Further significant reduction  of the required 
energy budget is still possible,  provided  
that the proton-synchrotron  radiation is responsible 
only for X-ray  emission, the radio and optical fluxes 
being related to other emission components, e.g. to 
the  synchrotron  radiation of electrons.  The dashed curve 
in  Fig.~3   corresponds to such  a scenario (model III). 
It is obtained  assuming the  following parameters:  
$B=3$ mG, $R=2$ kpc, $E_0=10^{18} \ \rm eV$,  
$\Gamma_0=2$, $L_{\rm p}=10^{45} \ \rm erg/s$.  
For these  parameters the proton synchrotron radiation 
explains the observed X-ray flux.  
However,  it appears  below the measured 
radio and optical fluxes,
because   neither the  synchrotron radiation 
nor the escape appear   sufficient  to cool effectively 
protons , and thus to steepen 
the original (acceleration) spectrum of protons at energies  
below $\leq 10^{17} \ \rm eV$ (dashed curve in Fig.~4).
A good  compensation for such a compromise is the relatively 
modest (for 3C 273) injection power of  
relativistic protons.  
 
\subsection{Synchrotron radiation of secondary electrons}

In   Fig.~4 we show the spectra of secondary electrons  
produced at interactions of protons with the ambient plasma 
and radiation fields.  The corresponding synchrotron
radiation of these electrons is shown in Fig. 3.    
The energy density of mm and sub-mm radiation, 
which is the most relevant band  of electromagnetic  
radiation for interactions of highest  energy protons, 
could be estimated  from Fig.~3 as  
$w_{\rm r} = \nu S_\nu (d/R)^2/c\approx 
10^{-13} \ \rm erg/cm^3$,  i.e. almost an order magnitude 
less than the energy density  of CMBR at the epoch  
$z=0.158$.  

The interactions of protons with CMBR  
do not  contribute significantly to the production 
of electrons with energy less than 100 TeV, i.e. 
to the production of electrons responsible for 
synchrotron radiation at radio to X-ray energies.  
The first generation electrons appear with energies 
exceeding $E_{\rm e} \geq  10^{15}  \rm eV$. On the 
other hand, the large magnetic field and small 
$\gamma-\gamma$ optical depths prevent an effective 
electromagnetic cascade development, which would allow 
production of lower energy electrons.   
As a result, the only signature of  $p \gamma$  
interactions seen Fig.~3  is the flattening of 
the spectrum of the electron synchrotron radiation 
at MeV/GeV energies (solid line),  caused  
by electrons produced through the  Bethe-Heitler  
pair production.

For relatively high gas density in the knot,
the interactions of accelerated protons with 
the thermal plasma contribute more effectively,
through  decays of $\pi^\pm$ mesons, in the 
production of relativistic electrons and 
positrons. Because of poorly known gas density 
in the jets,  the calculations of secondary ``$pp$'' 
electrons contain large  uncertainties. On the other hand, 
an important information  about the gas density  
could be provided by synchrotron  radiation of 
these electrons; the latter  obviously should not exceed 
the observed fluxes at radio, optical and X-ray bands. 
For the knots in 3C 273, the most informative  upper 
limits are contained  in the  very low radio fluxes 
at GHz frequencies.   The curves 
in Fig.~3  correspond to the gas density  
$n=10^{-5}  \ \rm cm^{-3}$ for the model I, 
and  $10^{-3}  \ \rm cm^{-3}$  for the models 
II and III.  It is seen that for the steep 
acceleration spectrum of protons $\propto E^{-2.4}$
the radio data only marginally  agree with the calculated 
synchrotron radio flux by secondary  electrons (solid line 
in Fig.~3), and  therefore  $n \leq 10^{-5}  \ \rm cm^{-3}$. 
Such a strong  constraint  is a result of 
very large amount of low, $\leq 10$ GeV  energy 
protons (solid curve in Fig.~4) - the
parents of secondary electrons producing 
synchrotron radio emission.  Since the gas density
in knots is likely to be significantly larger,  
this constraint could be considered as an independent 
argument against the  steep proton acceleration 
spectrum assumed in the the model I.  

The models II and III  with flat  $E^{-2}$  type 
acceleration spectrum of protons allow much higher,
respectively   $10^{-3}$, and   $10^{-2}  \ \rm cm^{-3}$
gas  densities in the knot. 
Because these numbers  are quite close to  the  densities 
expected in the environments of large scale extragalactic  
jets, it seems an attractive idea of referring a fraction,
or even the entire synchrotron spectrum of the knot 
to the synchrotron radiation of secondary  
``$pp$''  electrons.  The production spectrum 
of these electrons  with energies $\geq 1$ GeV 
coincides with the power-law  spectrum of protons, 
thus it is quite easy to guess the power-law 
index of acceleration spectrum of parent protons,  
taking  into account that  for the  magnetic 
field in the knot $B \sim $1 mG  and for the jet's 
age  $\geq 10^6$ yr,  the radio electrons effectively 
cool through the synchrotron radiation. Namely, 
$\Gamma_0=2 \ \alpha_{\rm r-x} \sim  1.5$.

The fluxes of  synchrotron  radiation of secondary  
``$pp$'' electrons  calculated in the  continuous
injection regime during last
$10^7 \ \rm yr$ are shown in Fig. 5 for two combinations
of model parameters:  
(a) $R=1$ kpc, $B=0.1$ mG, $\eta=1$; and
(b) $R=0.25$ kpc, $B=0.03$ mG, $\eta=10$.
In  both cases the same ``power-law  with 
exponential cutoff'' acceleration spectrum 
for protons is assumed 
with $\Gamma_0=1.5$ and  $E_0=10^{18} \ \rm eV$.
These parameters imply essentially different escape 
times  of protons with corresponding  breaks at 
$E_{\rm b} \simeq  3.5 \times  10^{16} \ \rm eV$  
and  $6.5 \times  10^{13} \ \rm eV$ for the 
cases (a) and (b), respectively. 
The resulting breaks in the synchrotron spectra of 
secondary electrons appear at 1 MeV and 1 eV,  respectively. 
It is seen the the case (a) explains  quite  
well the  observed nonthermal spectrum  from  
radio to X-rays, while the  case (b) can explain 
only radio and optical fluxes. Note that  for chosen 
parameters the highest energy protons quickly escape 
the knot. This effect, combined with the low magnetic field, 
results in a dramatic drop of emissivity 
of the proton synchrotron radiation  (dashed line in Fig. 5).

\begin{figure}
\epsfxsize=8.0 cm
\centerline{\epsffile[75 427 500 727]{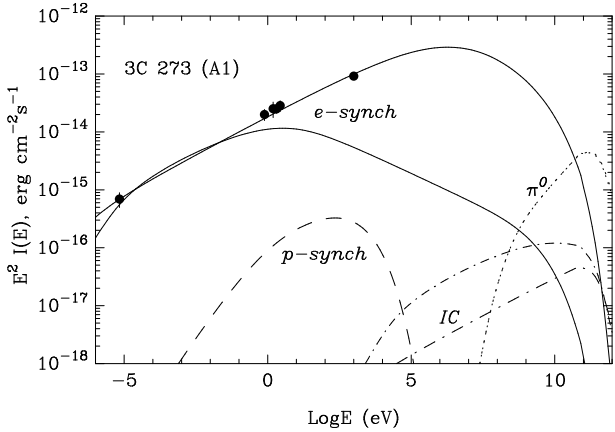}}
\caption{Broad-band nonthermal emission  of 
the knot A1 produced  directly by accelerated 
protons  via the  synchrotron radiation (dashed lines) 
and through the $pp \rightarrow \pi^0 \rightarrow \gamma$
channel (dotted lines), and  secondary electrons 
through the synchrotron (solid lines) 
and inverse Compton  (dot-dashed lines) 
radiation. The heavy lines correspond to the case 
of slow escape (model I) and thin lines correspond to 
fast escape (model II).  The  model parameters 
are  discussed in the text.
The synchrotron  radiation of protons for 
the  model II is too weak and is beyond the  
figure frames. The contributions of 
$\pi^0$-decay $\gamma$-rays for the models I and II  
are almost identical, and therefore cannot be distinguished. 
}
\end{figure}
 
The synchrotron radiation  by  secondary 
``$pp$'' electrons in radiogalaxies and AGN  
is not a new idea.    The basic  problem of this 
hypothesis -  the deficit of an adequate target material  -  
is clearly recognized since early 60's 
(see e.g.  Burbidge, Burbidge and Sandage 1963).
Our study of the A1 knot in 3C 273 faces  the 
same problem. In order to match the absolute 
fluxes of the observed nonthermal radiation from 
knot A1,  we must assume a very large product of 
the gas  density  and the proton  injection power,
$n  L_{\rm p}=8.5 \times  10^{46} \ \rm erg/s \ cm^3$.
Thus for any reasonable acceleration rate of protons 
$L_{\rm p} \leq 10^{47} \  \rm erg/s$,  the plasma density
in the knot should be close to  $1 \ \rm cm^{-3}$. 
Such high densities  are not supported, however,
by depolarization studies  of the radio emission 
(see e.g. R\"oser et al.  2000).  In principle,  this  
problem could be essentially softened if 
the knot is moving towards  the observer with a 
relativistic Doppler factor $\gg 1$.

\subsection{Relaxed model parameters due to 
relativistic beaming} 
Recently, Sambruna et al. (2001)
invoked the relativistic beaming effect to explain 
the X-ray emission from the  knots in  3C~273 
by the inverse Compton scattering of low energy electrons,
the main target photon field being the 2.7 K CMBR.
While this model can successfully explains the diffuse character 
of X-ray emission (the X-ray emitting electrons have 
long cooling times), it must assume a second, synchrotron
component  of radio to optical emission. 
Note  that the electrons responsible
for optical radiation should have very large energies 
of about 1 TeV, and correspondingly  cool on  very 
short times scales.  Therefore they still require 
{\em in situ} acceleration  (or re-acceleration).       

Obviously, the hypothesis of relativistic 
beaming has a universally ``positive'' 
effect for all radiation models. In particular 
it helps to reduce significantly the energy 
requirements to both  {\em proton synchrotron} 
and the  {\em secondary-electrons synchrotron}
models discussed above. 
Indeed, if the jet maintains the relativistic speed
up to kpc scales, for the Doppler factor of about  
$\delta_{\rm j} \sim 5$ (estimated for the inner jet 
of 3C~273; see e.g. Abraham \& Romero 1999),   
these  requirements become so relaxed 
that we can successfully explain 
the entire  radiation  of the knot A1 
merely by the proton  synchrotron radiation.  
At the same time,  the relativistic beaming makes 
the interpretation of the nonthermal emission of the knot 
by secondary ``$pp$'' electrons an  
attractive alternative with a quite 
reasonable  product of the gas density, 
the jet lifetime , and  the proton acceleration rate  
$(n/{\rm 0.1 \  cm^{-3}}) \ (\Delta T /{10^7 
\ \rm yr})\ (L_{\rm p}/{10^{45} \ \rm erg/s}) \sim 1$.

\subsection{Gamma-ray emissivity of the knots}

The quasar 3C 273 is identified as  powerful 
$\gamma$-ray emitter (Lichti et al. 1995) 
with a peak in the  spectral energy distribution at  
1-10 MeV at the level of  
$\simeq 4 \times  10^{-10} \ \rm erg/cm^2 s$;
the energy flux  at 1 GeV is about  
$\simeq 2 \times  10^{-11} \ \rm erg/cm^2 s$. 
These fluxes exceed by several orders of magnitudes 
the flux of the proton synchrotron radiation (see Fig. 3).
Below we briefly discuss two other potential 
mechanisms of $\gamma$-radiation related to 
interactions of relativistic protons.   

Besides the cooling through synchrotron radiation, 
the secondary electrons release their energy 
also through inverse Compton  radiation which 
may extend to  very high energies. However, for 
the parameters favorable  for the proton synchrotron  
radiation,  in particular for the magnetic field 
$B \geq$ 1 mG,  the contribution of the IC component  is 
negligible,  and therefore do not appear in Fig. 3. 
For smaller magnetic field, $B \leq 0.1$ mG,  
the IC fluxes are  higher (Fig. 5), but still well 
below the sensitivity of $\gamma$-ray detectors.  
This is also true for 
``direct'' $\pi^0$-decay $\gamma$-rays produced 
at $pp$ interactions. The fluxes of this component  
are shown in Fig.~3 and Fig.~5.
Note that the sharp drops of $\gamma$-ray 
fluxes above 100 GeV is caused by intergalactic 
photon-photon absorption. In calculations  presented  
in Fig. 3 and 5, we adopted  one of the  recent models 
of the diffuse extragalactic background of Primack et al. 
(2001) which satisfactorily describes the observational 
data at near infrared and optical wavelengths - the most 
important band from the point of view of intergalactic 
absorption of $\gamma$-rays above 100 GeV. 

Although the flux of $\pi^0$-decay $\gamma$-rays 
is proportional to the ambient gas density $n$ 
and the total amount of accumulated 
relativistic protons,
$W_{\rm p} \simeq \Delta t  L_{\rm p}$,   
it cannot be arbitrarily  increased assuming  larger 
values for the product $n  \Delta t  L_{\rm p}$. 
The  $\pi^0$-decay $\gamma$-rays are 
tightly coupled  with the synchrotron radiation 
of secondary  ($\pi^\pm$-decay) electrons. Therefore 
the fluxes of   $\pi^0$-decay $\gamma$-rays are 
(unavoidably) limited by the energy fluxes of 
synchrotron radiation of secondary electrons at  
low frequencies.  In particular, since  
both  GeV $\gamma$-rays and GHz 
radio photons are initiated by the same (10 to 100 GeV) 
primary protons, the flux of $\gamma$-rays around 
1 GeV cannot  exceed  $10^{-15} \ \rm erg/cm^2 s$,  
as it is clearly seen in Fig. 3 and Fig. 5. Obviously this 
upper limit  is insensitive to the relativistic beaming 
effects. This excludes any chance to explain 
the observed high energy $\gamma$-radiation from 3C~273 
by $pp$ interactions in the the large-scale jet. 
The observed MeV/GeV $\gamma$-radiation 
originates, most probably, in the inner jet.    

\section{Runaway protons outside jets}

Beyond 100 Mpc, the Universe is opaque for 
extremely high  energy (EHE) cosmic rays  
with energy $E \geq 10^{20} \, \rm eV$
(see e.g. Cronin 1999). Fortunately, observations 
of the electromagnetic radiation  associated with 
these particles can compensate, at least partly, 
this limitation, allowing  extension of the (indirect) 
study of EHE cosmic rays  to cosmological scales.

The results of previous sections show that  
an essential  fraction of the nonthermal energy 
of the jet released  in the knots and  hot spots in the  
form of relativistic protons, eventually 
escapes  the jet; the kpc scales of these 
structures are not sufficient for effective confinement 
of particles with $E \geq  10^{19} \ \rm eV$.     
Thus the nonthermal radiation of the jet does not tell 
us much about the most energetic particles accelerated 
in the  knots and hot spots.
Such a crucial information can be recovered by 
observing characteristic nonthermal radiation of 
the cluster environments  harboring the runaway particles.
The spatial and spectral characteristics of 
nonthermal emission  initiated by interactions of 
protons with the CMBR and the  ambient gas, generally
depend on both the spectrum of runaway protons and the 
properties of the local environment.
Below I  will discuss two different cases, 
assuming that (1) the source of EHE 
protons is  surrounded by  a  rich galaxy 
cluster with B-field 
exceeding $1 \ \rm \mu G$, and  (2) the EHE source  
is located in a low magnetic field  environment.

\subsection{Radiation initiated in rich galaxy clusters}

It is believed that a non-negligible fraction of the 
energy of clusters of galaxies is contained in non-thermal 
forms, in particular in the form of relativistic particles 
and magnetic fields.  While  this  conclusion concerning the 
energy budget of relativistic particles is still based on 
theoretical and phenomenological arguments comprehensively 
discussed 30-years ago  by  Brecher \&  Burbidge (1972), 
there is growing  evidence that many rich galaxy clusters 
have magnetic field strengths at the microgauss or even higher 
level (Kronberg 2001).  Remarkably, this concerns  not only
the rich galaxy clusters with strong cooling-flows, 
but also   the  ``normal'', non-cooling-flow clusters
(Clarke et al. 2001).  The estimated  magnetic field 
of about  5~$\mu \rm G$ with energy density 
$w_{\rm B}=B^2/8 \pi \approx 0.6  \ \rm eV/cm^3$ 
in the inner $R \sim 0.5$  Mpc sphere ( Clarke et al. 2001),  
would result, most probably, in relatively slow diffusion 
of cosmic rays in the intracluster medium, and 
thus would provide their effective confinement  and 
accumulation  in the galaxy clusters (V\"olk et al. 1996, 
Berezinsky et al. 1997, Blasi \& Colafrancesco 1999).  

The energy-dependent diffusion  coefficient, together with the 
initial spectrum of particles injected in the intracluster 
medium,  determine the  spectrum and the total energy content 
of protons established  in the cluster over lifetimes of about   
$t \sim 1/H_0 \sim 10^{10} \ \rm yr$.  
In Figs. 6a and 6b  we show two examples of  
contemporary proton spectra obtained within  the  
``leaky-box''  type approximation of particle propagation  
(see e.g. Ginzburg \& Syrovatskii 1964)  
for the following model parameters:

\vspace{2mm}

\noindent
{\bf (a)}  continuous injection of relativistic protons
with a constant rate $L_{\rm p}=10^{45} \ \rm erg/s$  
during $\Delta t=10^{10} \ \rm yr$;  
acceleration spectrum of protons with
power-law index $\Gamma_{\rm p}=2$ 
and  exponential cutoff at $E_0=10^{20} \ \rm eV$; 
intracluster magnetic field  $B=3 \ \rm \mu G$   
and  gas density $n=10^{-3} \ \rm cm^{-3}$ within 
the central $R=0.5 \ \rm Mpc$ region; escape time of protons 
from the  high magnetic field region 
$\tau_{\rm esc}=5  \times 10^7 E_{19}^{-1/2}  \ \rm yr$;

\vspace{2mm}

\noindent
{\bf (b)}  $L_{\rm p}=3 \times 10^{46} \ \rm erg/s$; 
$\Delta t=10^{8} \ \rm yr$;  
$\Gamma_{\rm p}=1.5$, $E_0=10^{20} \ \rm eV$,  
$R=0.5 \ \rm Mpc$,  $B=5 \ \rm \mu G$, 
$n=10^{-3} \ \rm cm^{-3}$,       
$\tau_{\rm esc}=2.0  \times 10^7 E_{19}^{-1/3}  \ \rm yr$. 

\vspace{2mm}

For the assumed magnetic field $B$ and the cluster 
radius $R$, the  chosen escape times correspond to 
diffusion coefficients given in the form 
$D(E)=D_0 E_{19}^{\beta}$  with  
$D_0$ and  $\beta$  compatible with the model parameters 
discussed in the literature (see e.g. V\"olk et al. 
1996, Berezinsky et al. 1997, Blasi \& Colafrancesco 1999). 
For the given diffusion coefficients, the particle  
escape becomes an important factor in formation of the 
proton  spectra above 
$E^\ast \propto (R^2/D_0 \ \Delta t)^{1/\beta} \sim  
3 \times 10^{14} \ \rm eV$
and $10^{17} \ \rm eV$,  for the cases  {\bf (a)} and 
{\bf (b)}, respectively. As a result, respectively  
$E_{\rm p}^{-2.5}$    and  $E_{\rm p}^{-1.83}$   
type proton spectra are   formed above these energies, 
before  approaching to the intrinsic (acceleration) 
cutoff at  $E \sim 10^{20} \ \rm eV$.

The cases {\bf (a)} and {\bf (b)} correspond to 
two essentially  different scenarios. 
The case {\bf (a)} could be treated as 
a quasi-continuous injection of protons into the cluster 
during its age  of about $10^{10} \ \rm yr$. In this case  
not only AGN jets, but also other sources 
of particle acceleration, e.g. large-scale 
(multi-hundred kpc) shock structures 
(Miniati et al. 2001) can contribute to  the high energy
proton production. The case {\bf (b)} correspond to the 
scenario when  the proton production is contributed by a 
single powerful AGN (see e.g. Ensslin et al. 1997)  
over its lifetime of about  $10^{8} \ \rm yr$. 
It should be noticed that,  because the escape time of
protons with energy $E \geq 10^{18} \, \rm eV$ is less than 
$10^{8} \ \rm yr$,  the relatively short time of 
operation of the accelerator has an impact only 
on the accumulated energy budget of lower energy protons.
In other words, the amount  of highest energy protons 
confined in the cluster is effectively determined by 
recent accelerator(s) operating since last 
$10^{7} - 10^{8}$ years.  
Also,  the weak energy dependence of the escape time
($\propto E^{-1/3}$) and  the  very hard proton spectrum 
($\propto E^{-1.5}$)\footnote
{Note that due to the confinement  of low-energy 
particles in the jet, the spectrum of protons injected 
in the intracluster medium should have a  low-energy 
cutoff. But since this effect does not have a  noticeable  
impact on the results, and in order to avoid  unnecessary 
increase of model parameters, we assume a single 
power-law injection spectrum.}, 
with the injection rate,  
$L_{\rm p}=3 \times 10^{46} \ \rm erg/s$,  provide 
much higher content of $\geq 10^{18}$ eV protons 
confined in the cluster for  the  
case ({\bf b}) compared with the case 
({\bf a}) (see Fig.~6). 
The assumed  proton acceleration rate is large,   
but still acceptable, given the fact that  
the total kinetic energy of electron-proton 
jets in powerful radio sources can significantly 
exceed $10^{47} \ \rm erg/s$ (Ghisellini \& Celotti 2001),  
provided  that the kinetic 
energy  power at the core is conserved along the jet up 
to $\geq 100$ kpc scales (Celotti \& Fabian 1993).      
The numerical calculations show that,  for the given 
duration and rate of proton injection, the total energy 
contents of protons accumulated 
in the cluster are
$W_{\rm p}=1.4 \times 10^{62} \ \rm erg$ and
$2.1 \times 10^{61} \ \rm erg$ for the cases {\bf (a)} and 
{\bf (b)}, respectively. These numbers do agree with 
general phenomenological considerations
discussed in the literature 
(Brecher and Burbidge  1972, V\"olk et al. 1996, 
Berezinsky et al. 1997,  Ensslin et al. 1997, 
Lieu et al. 1999). 

%
\begin{figure}
\vspace{10. cm}
\includegraphics{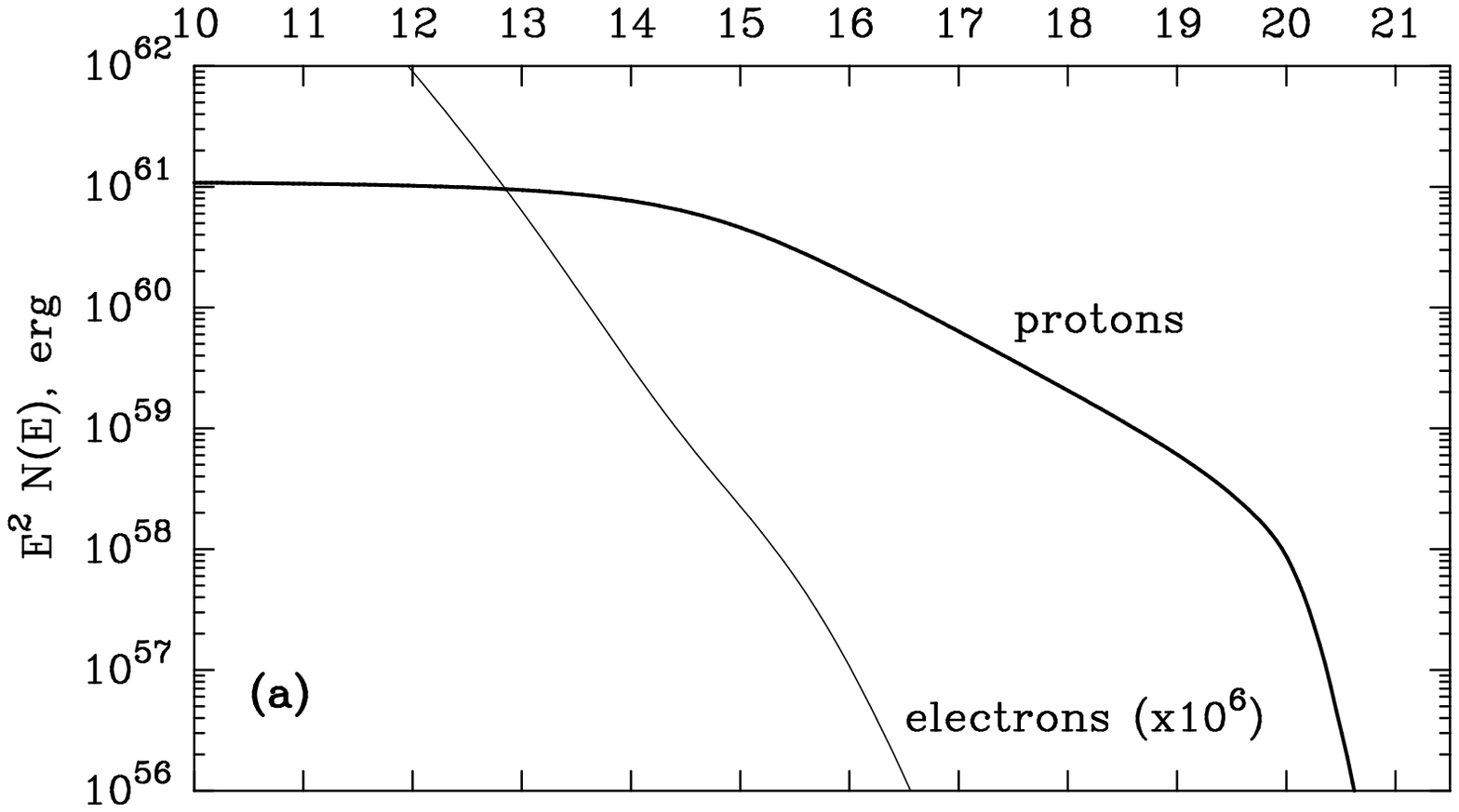}
\includegraphics{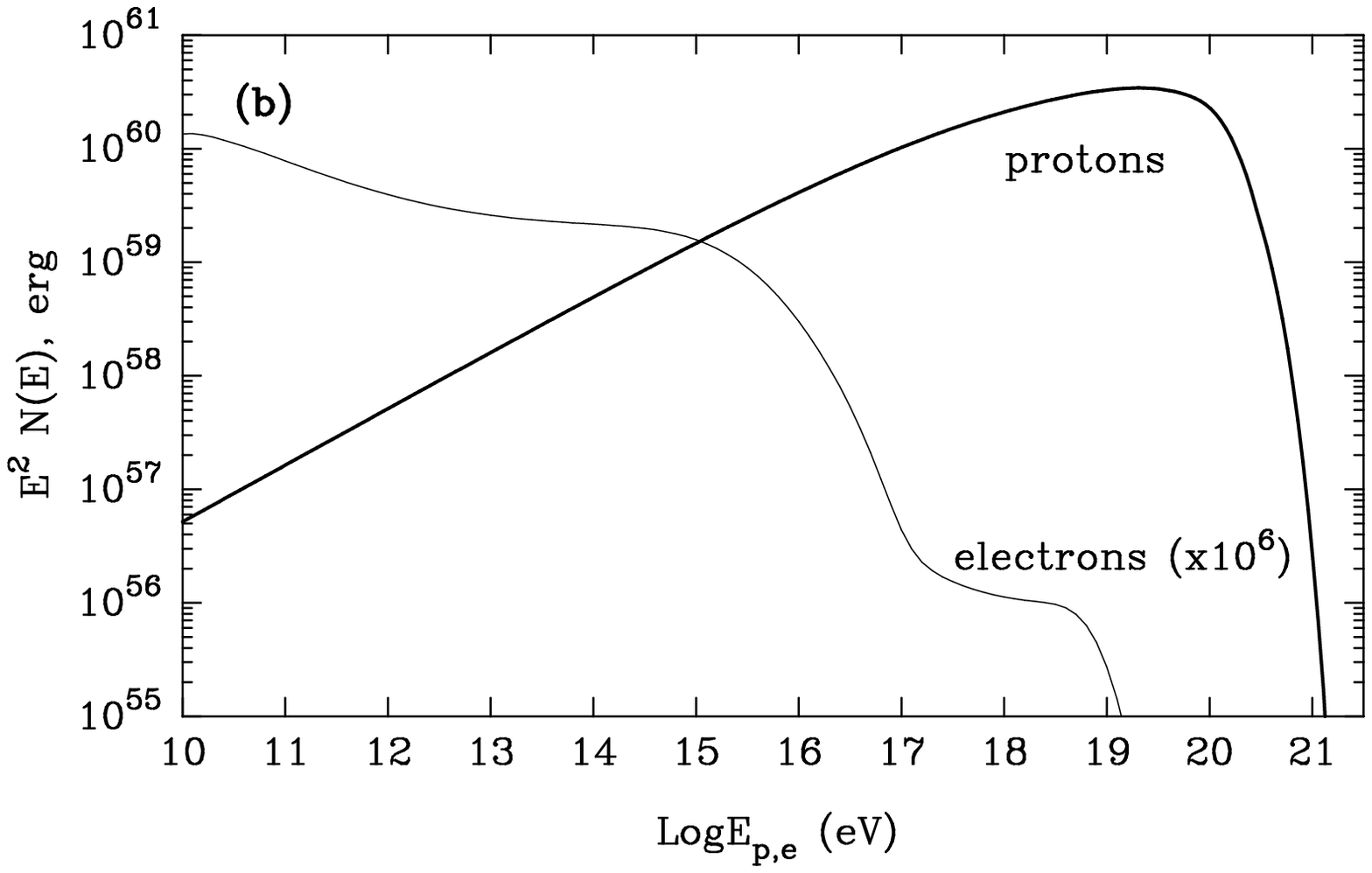}
\caption{The total energy content of protons and 
secondary electrons (multiplied by $10^6$) inside 
a galaxy cluster  calculated for scenarios {\bf (a)} and 
{\bf (b)} (see Sec. 5.1 for the assumed model parameters).  
}
\end{figure}

In Fig.~6 we show  the spectra of  secondary electrons
produced at interactions of relativistic  protons with
the intracluster gas for $n=10^{-3} \ \rm cm^{-3}$,
and with 2.7 K CMBR. These spectra  consist of three 
components:  
(i) electrons  from  $pp$ interactions, 
(ii) electrons from $p \gamma$ interactions 
due to the Bethe-Heitler pair production 
and (iii) electrons from the 
photo-meson processes. Due to the radiative cooling,
which  in the case of electrons strongly dominates 
over escape losses at  all energies, 
the first population electrons 
have a power-law spectrum 
$\propto E_{\rm e}^{-(\Gamma_{\rm p}+1)}$, while 
the  spectra of the second and third electron 
populations are almost independent of the 
spectrum of primary protons, namely   
both cooled electron populations have standard 
$E^{-2}$ type spectra with high-energy 
cutoffs at energies $\sim 10^{15} \ \rm eV$ and 
$\sim 10^{19} \ \rm eV$, respectively. 
While in the case {\bf (a)} the second and third ($ p \gamma$)  
electron components are strongly suppressed (Fig.~6a),  
in the case  {\bf (b)} the features of all three 
components are clearly seen in Fig.~6b.        

Although the energy content in secondary electrons
is very small compared with the total energy budget,
approximately half of the  nonthermal energy of the 
cluster is eventually radiated  through these 
secondary electrons. It is remarkable, that 
unlike primary (directly accelerated) electrons, which
because of  short lifetimes   
are concentrated
in the proximity of their  accelerators,  
the secondary electrons are homogeneously  
distributed over the entire cluster, and therefore their
radiation has an extended (diffuse) character.                     

The production rates of nonthermal radiation 
consisting of 5 components -- (1) synchrotron 
(marked as {\em e-synch}) and (2) inverse Compton 
({\em IC})  photons  emitted 
by secondary electrons,  (3) synchrotron radiation of protons 
({\em p-synch}),  and $\pi^0$-decay $\gamma$-rays from 
(4) proton-proton  ($pp$) and  (5) proton-photon 
($p\gamma$) interactions -- are shown in Figs.~7a and 7b.
It is seen that the spectral energy distributions (SED) 
of radiation characterizing  the {\bf (a)} and {\bf (b)} 
scenarios are essentially different.

In the case {\bf (a)} the nonthermal radiation is 
mainly contributed, directly or via secondary electrons,  
by  $pp$ interactions . The importance of these interactions 
in the clusters of galaxies has been often discussed in the
literature (Vestrand 1982;  Dermer \& Rephaeli 1988;
Dar \& Shaviv 1995;  V\"olk et al. 1996; 
Berezinsky et al. 1997;  Ensslin et al. 1997;
Atoyan \& V\"olk 2000). 
Since the  $pp$ interaction timescales  
in the galaxy clusters with  
$n \leq 10^{-3} \, \rm cm^{-3}$ 
exceed the source ages of about $10^{10} \ \rm yr$, 
the absolute fluxes of radiation are 
proportional to the product $n L_{\rm p} \Delta t$.
For the assumed index of accelerated protons 
$\Gamma_0=2$, approximately the same fraction 
of the proton kinetic energy is released in 
$\pi^0$-decay $\gamma$-rays and  $\pi^\pm$-decay 
electrons  and positrons. On the other hand, the 
assumed magnetic field  $B=3 \ \mu \rm G$   
implies an energy density close to the density of 
2.7 K CMBR, therefore equal fractions of 
the electron energy  are released  through the 
synchrotron and inverse  Compton channels. This  results 
in the flat  overall SED  over a very broad frequency  
range  from radio to multi-TeV $\gamma$-rays (Fig. 7a).  
The energy domain below 1 keV is due to   synchrotron  radiation 
of electrons,  while the interval between X-rays   and low energy 
($\leq 100$ MeV)    $\gamma$-rays is contributed  by inverse 
Compton  mechanism.  At higher energies  the  radiation is 
dominated by $\pi^0$-decay $\gamma$-rays produced at $pp$
interactions. The local maximum at 
$10^{19}- 10^{20} \,\rm eV$ in Fig.~7a, 
due to decays of $\pi^0$-photomesons, 
is relatively weak because of effective escape 
of protons from the cluster (see Fig.~6a)  
       
%
\begin{figure}
\vspace{10. cm}
\includegraphics{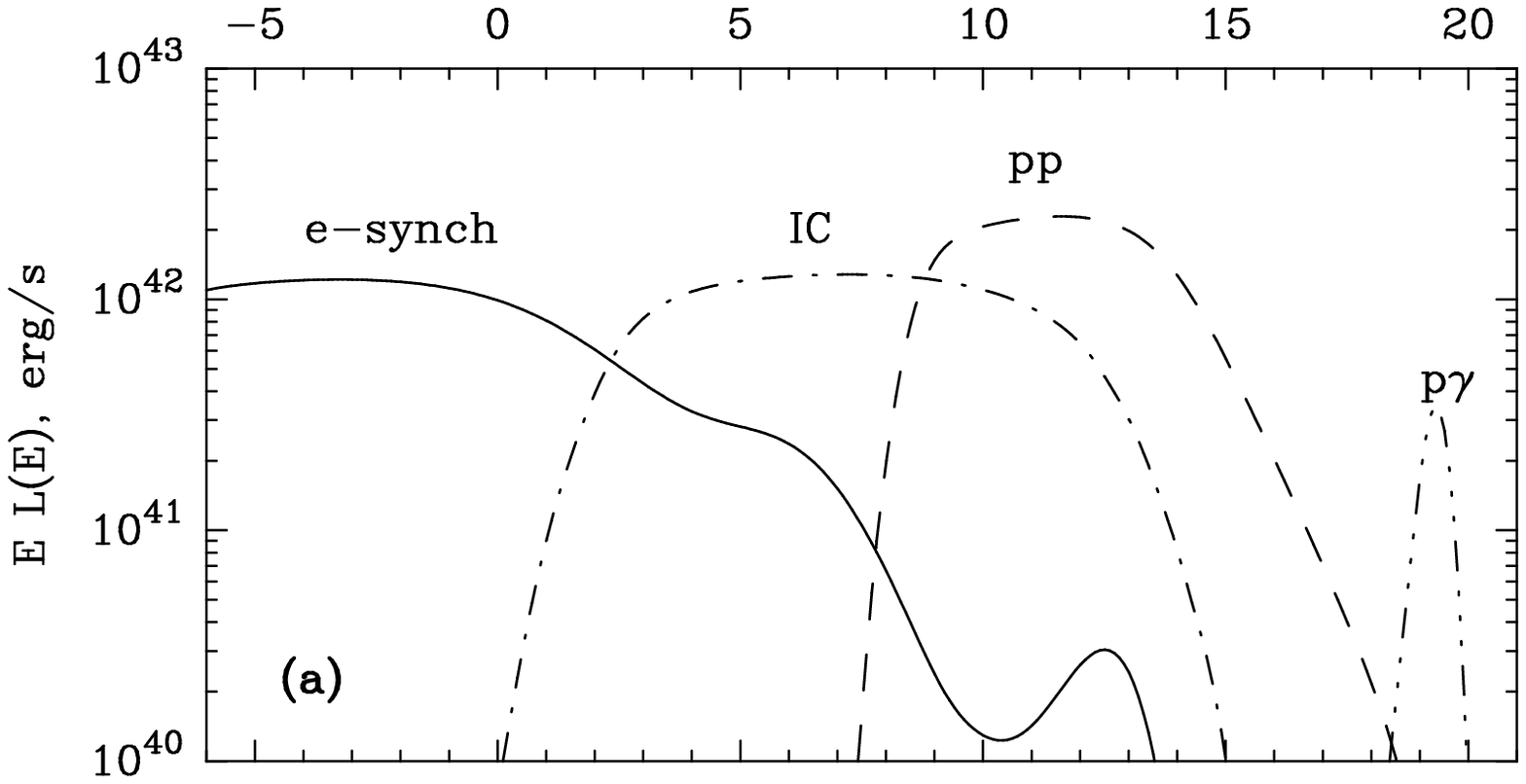}
\includegraphics{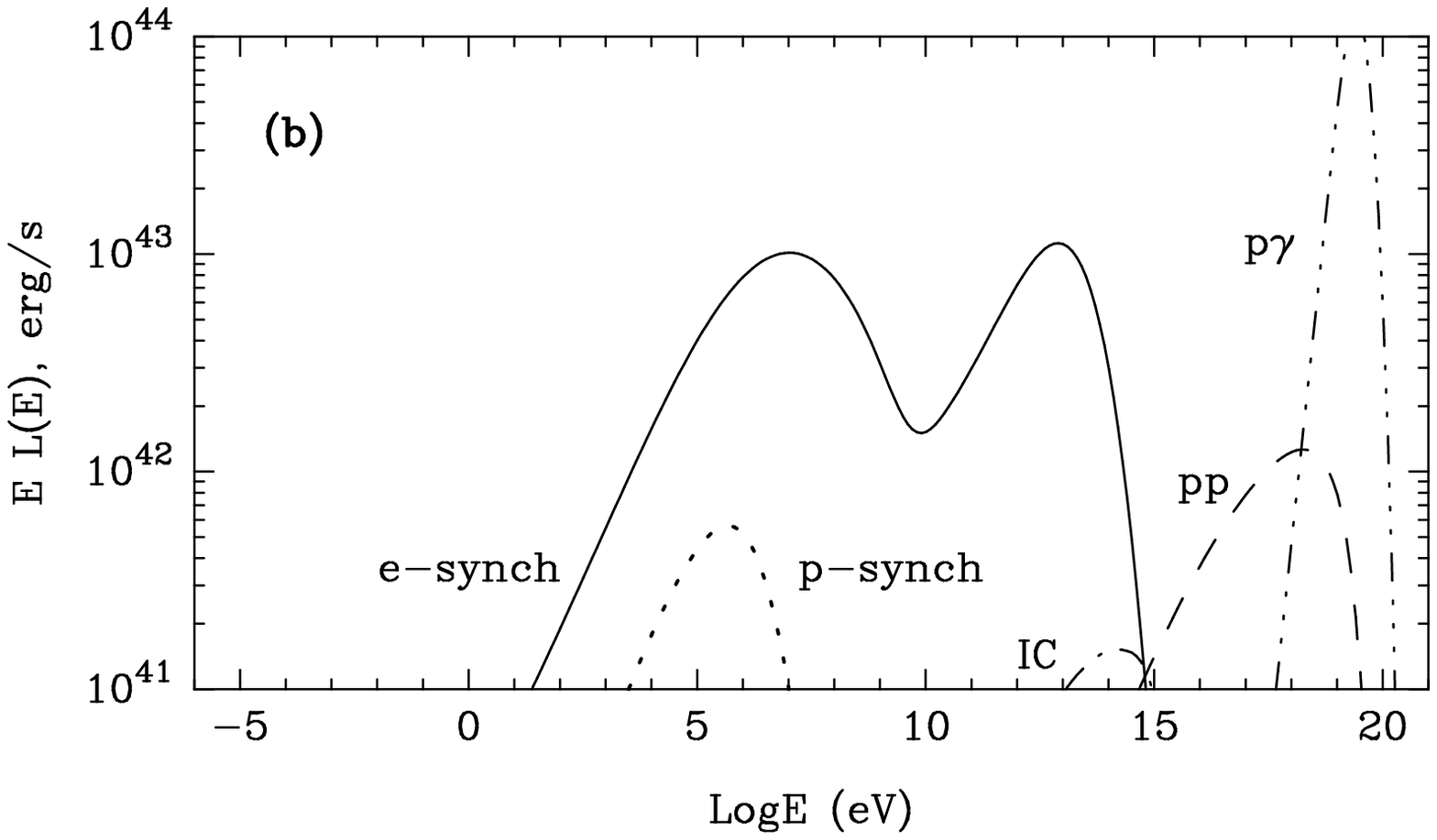}
\caption{Broad-band luminosities of 
nonthermal radiation initiated by protons 
in a galaxy cluster. 
The calculations correspond to  
two different scenarios of proton injection 
into the cluster (see the text for details). }
\end{figure}

For harder  spectra of accelerated protons, e.g. 
with power-law index $\Gamma_{\rm p}=1.5$ the SED
is strongly dominated by synchrotron radiation of secondary 
$p \gamma$ electrons  with two prominent peaks at MeV
and  TeV energies\footnote{Note that the synchrotron 
radiation by primary (directly accelerated) electrons 
has an unavoidable,  self-regulated cutoff below 
$\frac{9}{4}  \alpha_{\rm f}^{-1} m_{\rm e} c^2 \eta^{-1}
\simeq 160$ MeV (Aharonian 2000), the latter being 
determined by the balance between the synchrotron 
loss rate and the maximum-possible  ($\eta=1$) 
acceleration rate.  Obviously in  the case 
of synchrotron  radiation of  secondary electrons, 
there is no intrinsic limit on $\gamma$-ray energy.}
(Fig. 7b) corresponding to the radiation by electrons  
from Bethe-Heitler pair production and photo-meson  
production processes, respectively. 
While at optical and radio frequencies the energy flux
decreases as ($\nu S_\nu \propto \nu^{0.5}$), at 
extremely  high energies $E \sim 10^{19}-10^{20} \ \rm eV$ 
the prominent ``$p \gamma$'' peak dominates 
over the entire SED (Fig.~7b). However,  this peak, 
as a part of the whole region of $\gamma$-rays  above 10 TeV
is not visible for the observer. Because of interactions 
with the diffuse extragalactic background radiation, these 
energetic  $\gamma$-rays  disappear during their  
passage from the source to the observer.

%
\begin{figure}
\vspace{9.5 cm}
\includegraphics{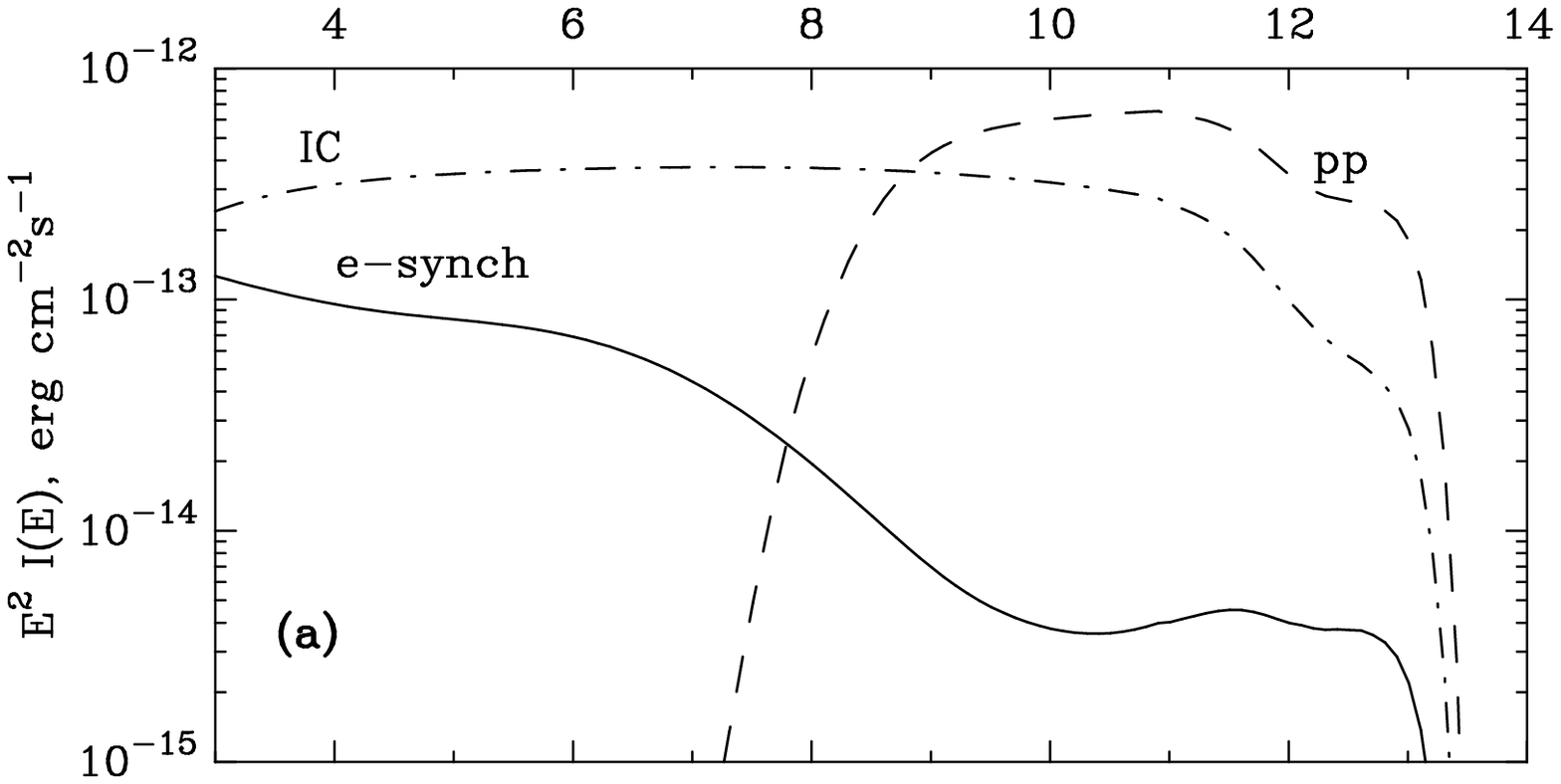}
\includegraphics{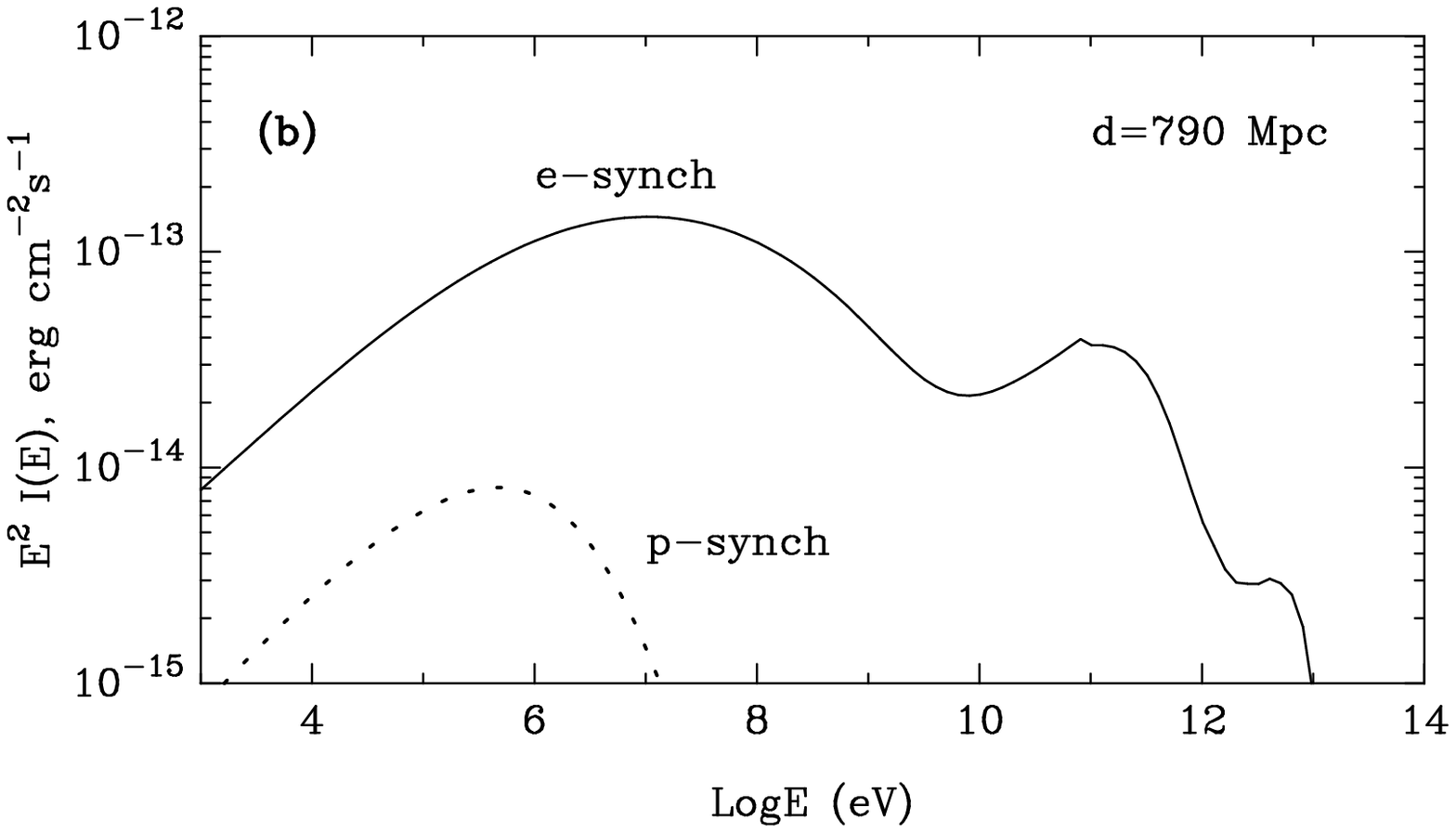}
\caption{Broad-band SED of nonthermal radiation of  
clusters corresponding to luminosities shown in Fig.~7a 
and Fig.~7b,  and assuming that the source is located 
at  $d=175$ Mpc ({\bf a}) and $790$ Mpc ({\bf b}), 
respectively (see the text for details and used model 
parameters.}
\end{figure}

\subsubsection{Detectability of gamma-rays}

In Fig. 8 we show the expected fluxes of radiation  
assuming that the sources with luminosities presented 
in Figs. 7a and 7b  are  located at small (like Pictor A) 
175 Mpc  (case {\bf a}) and large (like 3C 273) 790 Mpc 
(case {\bf b}) distances.  These are 
two specific examples chosen to have  a 
feeling for the detectability  of 
radiation at different  wavelengths.  
For the given source luminosities,  
the fluxes of  high energy $\gamma$-rays    
are determined not only by the distance 
to the source ($F_\gamma \propto 1/d^2$),
but also by the intergalactic photon-photon absorption.   
Note that the mean free path of $\gamma$-rays above 
100~TeV is less than 
1 Mpc, so  strictly speaking we should include in 
calculations the radiation of next generation 
electrons produced {\rm inside} the cluster.  
However, since the $\gamma$-ray 
luminosities  sharply drop above 100 TeV (see Fig.~7), 
we can ignore these (3rd generation) photons without 
a significant impact on the accuracy of 
calculations.  Such an approximation, however, 
could be inappropriate  for $\gamma$-rays with energy   
$\geq 10^{19}$ eV.  The mean free paths of these 
$\gamma$-rays, interacting with CMBR 
and extragalactic radio background photons, 
contain large uncertainties because of the lack 
of reliable information about the radio background 
at MHz frequencies (Berezinsky 1970, 
Aharonian et al. 1992, Protheroe \& Biermann 1996,
Coppi \& Aharonian 1997). Nevertheless it is 
likely that the  extragalactic photon fields 
cannot prevent  the $\geq 10^{19} \, \rm eV$
$\gamma$-rays to travel freely over distances 
more than 1 Mpc. Actually the interaction of the 
highest energy  $\gamma$-rays with 
the cluster's own  synchrotron radio 
emission could be more important process. 
It is easy to show that the $10^{19} \, \rm eV$
$\gamma$-rays can escape the central $\sim 1$~Mpc 
region of the cluster if the radio luminosity 
at frequencies $\sim 10 \, \rm MHz$
does not exceed $10^{42} \, \rm erg/s$. For comparison,
the 10 MHz radio luminosity of the 
Coma cluster  is about  $10^{40} \, \rm erg/s$, but in most 
powerful clusters it could be significantly higher.  
For example, the power-law extrapolation (with spectral 
index 1.6) of  the recently reported  radio flux from the 
giant halo in Abell~2163 ($z=0.203$) at 20 cm 
$S_\nu \simeq 0.15 \, \rm Jy$, to lower frequencies 
(Feretti et al. 2001) gives for the luminosity of the 
cluster $\simeq 5 \times 10^{42} \, \rm erg/s$ at 10 MHz. 
This implies  that the $\geq 10^{19} \, \rm eV$ 
$\gamma$-rays are absorbed inside this cluster.   
In such cases we have to  include in the overall SED  
the synchrotron radiation of the secondary electrons. 
This would simply result in  the linear increase 
(by a factor  of two) of the flux of $\geq 1$ TeV 
$\gamma$-rays (typical energy of synchrotron photons 
produced by $10^{19} \, \rm eV$ electrons in the magnetic field 
$B \geq 1  \, \mu \rm G$) shown in Fig.~8b. Even so, 
it would be difficult to detect this radiation due to 
the heavy intergalactic absorption of $\geq 0.1 \, \rm TeV$
$\gamma$-rays 
arriving from cosmologically distant ($d \gg 100$~Mpc) 
sources  (see e.g. Stecker et al. 1992).  
The chances to detect the 
``echo'' of the original $\geq 10^{19} \, \rm eV$ $\gamma$-rays 
would be dramatically increased if the free path of these
photons exceeds 1 Mpc. In this  case the photon-photon
interactions take place predominantly outside the cluster
where the intergalactic magnetic field $B_{\rm IG}$
is significantly smaller (most probably, 
$B_{\rm IG}  \leq 10^{-8} \, \rm G$). Then we 
should expect an extended synchrotron emission from the 
pair-produced (and radiatively cooled) 
electrons with hard differential spectrum 
$\propto E^{-1.5}$ up to the maximum energy 
in the SED expected at 
\begin{equation}
E_{\rm max} \sim 5 (B_{\rm IG} /10^{-9} \ \rm G) 
\ (E_{\rm e}/10^{19} \ \rm eV)^2 \ \rm GeV \ .
\end{equation}
The $\gamma$-ray spectrum beyond  $E_{\rm max}$
depends on the shape of the proton spectrum in the region 
of the cutoff $E_0$, but in any case it is smoother  
(a basic feature of the 
synchrotron radiation)  than the proton spectrum
in the  corresponding region above  $E_0$.
The energy flux of radiation which peaks at 
$E_{\rm max}$ can be easily estimated from 
Fig.~7b, assuming  that almost the entire 
$10^{44} \, \rm erg/s$ luminosity in the primary 
$E \sim 10^{19} - 10^{20} \, \rm eV$ $\gamma$-rays 
is re-radiated outside the cluster 
in the form of secondary synchrotron photons.
Namely, we should expect an energy 
flux at the level 
$F_\gamma \sim 10^{-12} (d/1 \, \rm Gpc)^{-2} \, \rm erg/cm^2 s$
contributed mainly by the region around $E_{\rm max}$.
It is interesting to note that although the 
secondary synchrotron radiation is produced
outside the cluster, the angular distribution 
of this radiation seen by the observer would essentially 
{\em coincide  with the size of the cluster}. Indeed,  
the ratio of the gyroradius to the
mean synchrotron interaction path of electrons,
which is estimated as  
$r_{\rm g}/\lambda_{\rm synch} \simeq
30 \ E_{19}^2 (B_{\rm IG}/10^{-9} \, \rm G)$, implies  
that the photo-produced electrons of energy 
$E_{\rm e} \sim 10^{19} \, \rm eV$ will radiate 
synchrotron photons  before they would 
significantly change their 
direction in the magnetic field 
$\geq 10^{-9} \, \rm G$, i.e. the  
synchrotron $\gamma$-rays produced 
outside the cluster will follow the 
direction of their ``grandparents'' -- 
$\pi^0$-decay $\gamma$-rays 
produced inside the cluster. 
Below  $E_{\rm max}$ 
the angular distribution of $\gamma$-rays could be 
quite different. In this energy region 
$\gamma$-rays produced by cooled 
electrons with steady state 
distribution $\propto E_{\rm e}^{-2}$,  
have hard power-law 
differential spectrum $\propto E^{-1.5}$. 
For $B_{\rm IG} \sim 10^{-9} \, \rm G$, 
the trajectories of $E \leq 10^{18} \ \rm eV$ 
electrons will be curved significantly before they radiate 
synchrotron $\gamma$-rays. Therefore,    
the size of the low energy $\gamma$-rays, 
$E \leq 100 \, \rm MeV$, produced by these electrons, 
is  determined by the 
mean free path of primary $\sim 10^{19} \, \rm eV$
$\gamma$-rays in 
the intergalactic photon field, and thus would exceed 
the angular size of the cluster.
The performance of the GLAST instrument 
(see e.g. Gehrels \& Michelson  1999)
is  perfectly suited to  both  
the expected fluxes  and the energy range of 
this  radiation even from  cosmologically 
distant ($d \geq 1 \, \rm Gpc$) sources.  Detection of 
such radiation would not only confirm the presence of 
highest energy protons in the clusters of galaxies, but 
also could provide unique information about the 
intergalactic magnetic fields  and the diffuse 
extragalactic radio background.

\subsubsection{Detectability of X-rays}    
 
The ``hadronic'' nonthermal radiation luminosities 
shown in Fig.~8 depend  on several model parameters 
like the magnetic field and the gas density of the  
intracluster medium, the total amount of protons 
confined  in the cluster, etc. The shape 
of radiation strongly  depends on the spectrum of 
injected protons, as well as on the  diffusion coefficient. 
However, since both the assumed ambient gas density and 
the proton injection rates are close to the maximum 
allowed  numbers,  the luminosities  shown in Fig.~8
hardly could be dramatically 
increased. If so,  the nonthermal X-ray 
fluxes of ``hadronic'  origin could be marginally 
detected only from relatively nearby clusters like Coma, 
Perseus, the cluster surrounding Cygnus A, etc. The case 
of the Coma cluster presents a special interest because 
of hard, most probably nonthermal X-radiation 
recently reported from this source (see e.g. Rephaeli 2001).  
The most natural  interpretation of this radiation  
by the inverse Compton scattering of radioelectrons 
requires an intracluster magnetic field 
$\sim 0.1 \, \rm \mu G$ (e.g. Atoyan \& V\"olk 2000), 
i.e. noticeably smaller than the 
the value of about several $\mu G$ deduced from 
Faraday rotation  measurements. Although this discrepancy 
perhaps should not be overemphasized (for  possible 
explanations see Rephaeli 2001, and Petrosian 2001), 
it is worth exploring other possibilities,  
in particular invoking  the radiation of  secondary 
electrons produced at interactions of EHE protons with the 
ambient gas or photon fields. 
 
Concerning the scenario ({\bf a}), Fig. 7a predicts a  
radio luminosity $\sim 10^{42} \, \rm erg/s$ which is by 
two orders of magnitude higher than the luminosity observed 
from Coma. Because of the assumed large magnetic 
field $B=3 \, \rm \mu G$, we face the same problem which 
arises in the  standard inverse Compton models. This  
imposes an upper limit on  the X-ray luminosity, 
$L_{\rm X} \leq 10^{40} \, \rm  erg/s$.    
However, this should not be treated as a 
robust constraint. In the   ``hadronic'' 
models this limit can be easily  removed  by assuming
harder (than $E^{-2}$) power-law proton spectrum,   
and/or a low-energy  spectral cutoff. Of course, 
this would automatically suppress also the inverts Compton 
component of radiation produced by low energy 
secondary electrons. In  
contrast to the conventional 
inverse Compton  models, in the ``hadronic''  
model we have an additional important channel 
for  production of X-rays - synchrotron 
radiation of very high energy (multi-TeV) secondary electrons. 
This is demonstrated in Fig.~7b (scenario ({\bf b})). 
Even so, for the distance to Coma $\sim 100$ Mpc, the 
expected  hard X-ray flux is below, by an order of magnitude,  
the reported 20-80 keV flux from Coma of about 
$2 \times 10^{-11} \, \rm erg/cm^2 s$. In the ``hadronic'' 
model the synchrotron X-ray luminosity slightly depends 
on the strength of the  magnetic field, thus in order 
to compensate this deficit,  we have to increase the 
proton injection rate by the same 
factor, i.e. up to $> 10^{47} \, \rm erg/s$. 
This uncomfortably high, at least for an object 
like Coma, acceleration power makes problematic, 
although cannot completely exclude, the ``hadronic'' 
origin of X-rays.  Decisive tests could be 
provided by new spectroscopic measurements (the 
hadronic model unambiguously predicts a 
hard X-ray spectrum  with photon index $\sim 1.5$
up to 1 MeV).  Also, the model predicts large  
$\gamma$-ray  fluxes both at MeV/GeV and TeV energies. 
These components have, however,  different origins. 
The radiation up to 10 GeV 
is produced by ``Bethe-Heitler'' electrons, while the TeV 
$\gamma$-rays are due to the electrons from $\pi^\pm$ decays.
While the TeV luminosity would be  
suppressed if the exponential cutoff in the 
proton spectrum occurs below  $10^{20} \, \rm eV$, 
without any impact on the X-ray flux, 
the MeV/GeV $\gamma$-rays are tightly connected with 
X-rays. This implies that we should expect 
similar fluxes in $\leq 100 \, \rm keV$ 
X-rays and $\geq 100 \, \rm MeV$ $\gamma$-rays, 
i.e. if the ``hadronic'' origin of X-rays is correct, then  
$F_\gamma(\geq 100 \, \rm MeV)
\sim 10^{-11} \, \rm erg/cm^2 s$. This 
hypothesis can be  easily checked by GLAST.   
    
Finally we note that  the detection of ``hadronic'' 
X-rays from extended regions of clusters of galaxies 
is a rather hard task, even for 
instruments like Chandra and XMM-Newton, especially 
because of the presence of high  local X-ray 
components like  the  thermal X-ray emission of 
the hot intracluster gas,  nonthermal X-rays  
due to inverse Compton radiation of  directly 
accelerated electrons, etc. Regardless of the details, 
the spectral band of  high energy $\gamma$-rays seems 
a more promising  window to explore the ``hadronic'' processes 
with GLAST   and, perhaps also with  forthcoming  100 GeV 
threshold  Cherenkov telescope arrays. 

\subsection{Very high energy radiation produced in  
low-magnetic-field environments}  

The characteristics of nonthermal radiation  of runaway 
protons in the case of a lack of strong cluster environment 
surrounding the EHE source,  are essentially  different  
from the radiation features described  in the previous section.  
The weak magnetic fields in such regions make 
faster the particle propagation, as well as  
prevent the dramatic synchrotron  cooling of highest 
energy (first generation) electrons. This allows 
an effective  development of relativistic electron-photon  
cascades triggered by interactions of runaway protons 
with  2.7 CMBR.  
The first stage of the cascade initiated by 
secondary   electrons and $\gamma$-rays  
interacting with the same  
2.7 K CMBR leads to formation of a 
standard $\gamma$-ray spectrum 
which can be  approximated as 
$dN_{\gamma}/dE \propto E^{-1.5}$ at $E \leq 10 \, \rm TeV$, and
$dN_{\gamma}/dE \propto E^{-1.75}$ at $E >  10 \, \rm TeV$ with 
a sharp cutoff at $E \sim 100 \, \rm TeV$. After the fast 
development in CMBR,   the cascade enters the second  
(slower) stage.  At this stage 
$E \leq 100 \, \rm TeV$ $\gamma$-rays produce $e^{\pm}$ pairs 
on the IR/O diffuse  background radiation,  while the Compton 
scattering  of  electrons is still dominated  by CMBR photons  
(Berezinsky et al. 1990;  Protheroe \& Stanev 1993; Coppi \& 
Aharonian 1997).   The second-stage  cascade, which actually 
consists of  2-3  interactions, shifts the spectrum to  lower 
(TeV and sub-TeV) energies, and  broadens  the  
angular distribution  of emission due to the deflections of 
$E < 100 \, \rm TeV$ electrons in the ambient magnetic field. 
Unfortunately, the  intergalactic magnetic fields 
and their fluctuations in very large ($\gg 1$~Mpc)  
scales remain highly unknown, which does  not allow us 
to make definite conclusions concerning the expected 
characteristics  of radiation. Nevertheless, depending on the 
strength  of intergalactic magnetic field,  
one  of the following $\gamma$-ray emission
components  can be predicted. 

\paragraph*{Very week magnetic field.}
For a very weak intergalactic field, 
$B \leq  10^{-15}$~G\footnote
{Although  quite speculative, such a large-scale 
intergalactic magnetic field  
cannot be {\it a priori} ruled out
(see e.g. Plaga 1995; Waxman \& Coppi 1996), 
in particular if an essential fraction of the Universe 
consists of  huge,  100 Mpc scale voids (Einasto 2001).},
the cascade radiation arrives, 
because of  almost rectilinear  
propagation of primary  protons and secondary  pairs, 
from  a direction  centered on 
the source.  In this case we  may  expect  a point-like source 
of radiation,  although the  $\gamma$-rays  are produced at 
distances  $\geq 10$ Mpc 
from the source. Indeed,  for the given energy  of a detected  
$\gamma$-ray photon  $E_{\rm TeV}=E/{1 \ \rm TeV}$,
the emission angle  is determined by the direction of  
electrons participating in the last  interaction, 
namely by  the  deflection in the magnetic field 
of the parent electron of energy 
$E_{\rm e}=(E/4 kT)^{1/2} m_{\rm e}c^2 
\simeq 17 E_{\rm TeV}^{1/2} \, \rm TeV$. 
The mean attenuation path of these electrons
in the CMBR is about 
$\Lambda_{\rm e} \simeq 0.02 E_{\rm TeV}^{-1/2} \, \rm Mpc$,
while the gyroradius is 
$r_{\rm g} \simeq  20 \ E_{\rm TeV}^{1/2}   (B_{\rm IG}/10^{-15} \, \rm G)^{-1} \, 
\rm Mpc$.
Correspondingly,  the emission angle  
$\theta(\epsilon) \sim \Lambda_{\rm e}/r_{\rm g}
\sim 10^{-3}  E_{\rm TeV}^{-1}  
\ (B_{\rm IG}/10^{-15} \, \rm G)$. Thus, for the intergalactic 
magnetic  field of about $10^{-15} \, \rm G$,  the cascade 
radiation at $E \geq 100 \, \rm GeV$  would be concentrated 
within  an angle of $1^{\circ}$.  Since 
for sources at distances between  100 Mpc and 1000 Mpc,
we expect hard cascade spectrum with a cutoff  
between  100 GeV and 1 TeV, an   
approximately half of the energy of  
$\geq 10^{20} \ \rm eV$ protons (completely lost at interaction 
with CMBR  over distances $\leq 100$ Mpc)  would be released  
in this energy interval,  and give a flux
\begin{equation}  
F_\gamma \sim 5 \times 10^{-12} 
\left(\frac{L_{\rm p}(\geq 10^{20} \, \rm eV)} 
{10^{45} \ \rm erg/s}\right) 
 \left(\frac {d}{1 \, \rm Gpc}\right)^{-2} \, \rm erg/cm^2 s  
\end{equation}   

Down to lower energies,  the energy flux   
decreases ($\propto E^{1/2}$),  and  
the angular size of emission increases ($\propto E$). 
Therefore searches for such an emission from directions 
of nonthermal extragalactic sources, in particular 
from AGN with  powerful X-ray jets, can be
done most effectively by 100 GeV threshold 
imaging atmospheric Cherenkov telescopes 
(e.g. Aharonian \& Akerlof 1997).   
  
\paragraph*{Intermediate magnetic field.}
If  $B_{\rm IG} \geq 10^{-12} \, \rm G$,
the cascade electrons  are promptly isotropized.
This leads to the formation of giant pair ``halos'' 
surrounding strong extragalactic TeV  sources 
(Aharonian et al. 1994).  The angular size of the 
extended $\gamma$-ray source depends on photon energy. 
For the  given energy of the  
detected photon $E_{\rm TeV}$, it is mainly determined
by the mean free path of previous generation 
$\gamma$-rays of energy  
$E^\prime \simeq 2  E_{\rm e} 
\simeq 34 E_{\rm TeV}^{1/2} \, \rm TeV$.
The free path of $E^\prime \sim 10-15 \, \rm TeV$ 
photons,  which are 
responsible for the  detected (last generation) 100 GeV 
cascade $\gamma$-rays,  presently is poorly known,
but,  probably,  it does not exceed 50 Mpc (see e.g. Primack 
et al. 2001). Thus the typical size of a 
100 GeV halo radiation surrounding 
the EHE source at a distance 1 Gpc would be  
$\leq 3^{\circ}$.  The  detection of pair halos presents a
difficult experimental task, compared, in particular, with 
the detection of  rectilinear cascade radiation discussed above. 
At the same time, as a compensation,  the 
EHE sources can  be revealed through their halo radiation
independent of the orientation of AGN jets.  It should be 
noticed in this regard that  only in the case of an isotropically 
emitting  source  the halo  will be centered on the source. 
If the relativistic outflow injecting EHE protons 
is directed away from the observer,
the center of the halo would be displaced by an angle 
comparable to the typical angular size of the halo. 
The radiation characteristics of halos 
{\em  initiated by  EHE protons}  primarily depend on the 
level of the  diffuse extragalactic background,  
first of all  at mid-  and  far- infrared wavelengths, but 
not on the  intergalactic  magnetic field,   
provided that the latter does not exceed  $10^{-9} \, \rm G$.  
    
\paragraph*{Large magnetic field.}
When considering  electromagnetic cascades 
initiated by EHE protons in the 
intergalactic medium with magnetic field  
$B_{\rm IG} \geq 10^{-9} \, \rm G$, we 
must distinguish between two populations 
of secondary electrons. 
The electrons originating from the Bethe-Heitler 
pair production process are produced with typical energies
$(m_{\rm e}/m_{\rm p}) E_{\rm p} \sim 10^{15}-10^{16} \ \rm eV$.
For magnetic field $\leq  10^{-7} \ \rm G$ they cool 
mainly through inverse Compton scattering, 
and thus  produce {\em faint}  $(e^+,e^-)$ halos  in a way 
discussed above.  Meanwhile,  
the electrons  originating from the photo-meson
production process, directly,  via $\pi^+$-decays,  
or through interactions of $\pi^0$-decay $\gamma$-rays 
with the extragalactic diffuse radio background,  
have much higher  energies, 
$\sim 1/10 E_{\rm p} \geq 10^{19}$ eV,  taking into 
account that  only $\geq 10^{20}$ eV protons interact effectively 
with CMBR.  Because of the Klein-Nishina effect, 
the energy losses of these electrons are dominated by 
the synchrotron radiation,  as long as  the ambient 
magnetic field exceeds   
$\sim 10^{-9} \ \rm G$ (Gould \& Rephaeli 1978).  
This prevents the cascade development, but instead provides  
another effective channel for production of  
high energy $\gamma$-rays. Almost the whole energy
of $\geq 10^{20} \, \rm eV$ protons is released, through
the synchrotron radiation of secondary electrons,
into the $\gamma$-rays with characteristic energy   
$\epsilon _{\rm max} \sim 50 (B/10^{-8} 
\ \rm G) (E_0/10^{20} \ \rm  eV)^2 \ \rm GeV$. 
Because the gyroradius of 
$10^{20} \, \rm eV$ protons in the magnetic field 
$B_{\rm IG} \geq 10^{-9} \, \rm G$ is comparable or less   
than their  mean $p \gamma$ interaction  path 
$\Lambda_{\rm p} \sim 100 \ \rm Mpc$, 
we should expect a diffuse radiation 
component emitted by huge intergalactic regions. 
This  diminishes the  chances for detection of this 
radiation component, 
especially from relatively nearby ($d \ll 1 \ \rm Gpc$) objects.
The situation is quite different  for cosmologically distant 
objects.  Because at distant cosmological epochs 
the CMBR was denser, $n_{\rm ph} \propto (1+z)^3$,
and hotter, $T_{\rm r} \propto 1+z$, the mean free path 
of protons $\Lambda_{\rm p}$  has a strong z-dependence.       
At energies $E_{\rm p} \leq 3 \times 10^{20} \ \rm eV$
it  can be approximated as 
$\Lambda_{\rm p} \simeq 5.2 (1+z)^3 
\exp[{3 \times 10^{20} \ \rm eV}/(1+z)E] \ \rm Mpc$  
(Berezinsky \& Grigoreva 1988).  For example,   
in  the  environments of quasars 3C~273 ($z=0.158$) and 
PKS~0637-752    ($z=0.651$), the mean free paths of 
$10^{20} \ \rm eV$ protons are 44.6 Mpc and 7.1 Mpc, 
respectively.  This implies that, if the X-ray emission 
from large scale jets of  these powerful  quasars has indeed 
proton-synchrotron origin, we may expect an
accompanying GeV $\gamma$-radiation component  
initiated by highest  energy run-away protons 
outside the jets,
but still  within  $\sim 3^{\circ}$ centered on  3C 273,
and  within 10 arcminutes centered on  PKS~0637-752. For some 
model parameters,  discussed in Sections 3 and 4, the  
expected  fluxes of this component of radiation would be   
sufficiently high to be detected by GLAST.  
 
\section{Summary} 

The current models of large scale AGN jets 
relate  the nonthermal X-ray emission of
distinct jet features, like the knots and hot spots, 
to the synchrotron or inverse Compton radiation of 
directly accelerated electrons. Both models however face 
serious problems. The inverse Compton or synchrotron-self 
Compton  models typically fail  on energetic grounds, unless
one assumes that the X-ray emitting regions are jet 
structures moving relativistically towards the observer. 
This is  a very promising approach  which seems to be applicable
for the jets in 3C 273 and  PKS 0637-752, but needs further  
inspections based on larger source statistics. 

While  the weakness of the inverse Compton models originates 
from the lack of sufficiently dense photon target fields,
the electron synchrotron  model  has  just an
opposite problem. It is  ``over-efficient'' in the sense that
the TeV electrons  due to severe radiative losses have 
very  short propagation lengths, and thus  do not allow 
formation of diffuse X-ray emission on kpc scales,
as it is observed by Chandra.        
A possible solution could be that the electron acceleration  
takes place throughout entire volume of a knot or a hot spot. 
However the operation of  huge, kpc size  accelerators 
seems to be a non-trivial 
theoretical challenge. The secondary origin of TeV  electrons,  
produced homogeneously in the knots  by relativistic 
protons due to  interactions  with the ambient gas, seems an 
interesting possibility.  But this hypothesis 
requires unacceptably large 
product of the gas density  and the proton acceleration 
rate, unless we assume that the X-ray emitting regions 
are relativistically moving jet structures  
with Doppler factors $\delta_{\rm j} \gg 1$.    

In the present work  I propose an alternative  
mechanism for X-ray emission  -  the synchrotron radiation
of extremely high energy protons with 
$E_{\rm p} \geq 10^{17} \, \rm eV$ accelerated in the  in  
large-scale  jet structures.  The idea advanced in the 
paper is that it is possible to construct a realistic  
model which allows effective cooling of protons via 
synchrotron radiation on quite comfortable timescales of 
about $10^7 - 10^8$~yr, i.e. on timescales which provide 
effective propagation of protons over the jet structures 
on kpc scales. This explains 
in a rather natural way not only  the diffuse  character of the 
observed  X-ray emission, but  also  the broad range of spectral 
indices  observed from different objects. Yet, the model
provides quite high radiation efficiencies if we allow 
relatively large magnetic fields in the  knots and  hot spots at the 
level  1 mG.  For example,  the required   
proton acceleration rates ranges from 
$10^{43}-10^{44}$ erg/s  in  the $25^{\prime \prime}$ 
knot in 3C 120  and in the west hot spot  of Pictor A,
to  $10^{45}-10^{46}$ erg/s in  the knots of 
powerful quasars 3C 273 and PKS 0637-752.
For relativistic jets  aligned with the line of sight 
these numbers can be  reduced by two or three orders of 
magnitude. 

An  essential fraction of the jet energy released in 
the form of extremely high energy protons eventually 
escapes the jet;   the kpc scales of knots and hot spots  
are not sufficient for effective  confinement of the most  
energetic particles with  $E_{\rm p} \geq 10^{19} \ \rm eV$.   
The runaway protons interacting  with the CMBR photons, 
as well as with the  ambient gas,    
initiate  non-negligible  nonthermal radiation components  in   
the cluster  environments  of AGN and radiogalaxies.
The signatures in the spatial and spectral distributions
of this emission, which extends from radio wavelengths 
to very high energy $\gamma$-rays, contain  unique  
information about both the 
highest energy protons accelerated in the AGN jets and about the 
the intergalactic magnetic  field in the  extended intergalactic 
environments around powerful AGN and radiogalaxies. 

\section*{ Acknowledgments}                            
                                                  
I wish to thank the participants of the   
``LSW/SFB 439 -- AGN'' seminar for the friendly 
discussion of the results presented in this paper. 
I am  grateful to Markos Georganopoulos and  John 
Kirk for helpful comments and discussions, 
and especially to Armen Atoyan for our fruitful 
collaboration and his important contribution
at the initial stage of this study.  
                
{}
\end{document}